\documentclass[%
 reprint,
 amsmath,amssymb,
 aps,
]{revtex4-2}

\usepackage{graphicx}
\usepackage{dcolumn}
\usepackage{bm,hyperref, mathtools}
\usepackage{natbib} 
\usepackage{tabularx} 
\usepackage{amsmath}  
\usepackage{color}

\newcommand{\hf}{\hat{F}_0}
\newcommand{\hv}{\hat{\mathbf{v}}}
\newcommand{\hdr}{\hat{D}_R}
\newcommand{\htt}{\hat{t}}

\newcommand{\ud}{\textrm{d}}

\newcommand{\kbT}{k_\mathrm{B}T}
\newcommand{\tlj}{t_\mathrm{LJ}}

\definecolor{darkgreen}{rgb}{0,0.5,0}
\definecolor{blueother}{rgb}{0,0.8,1.0}

\begin{document}

\title{How boundary interactions dominate emergent driving of passive probes in active matter}

\author{Jeanine Shea$^1$, Gerhard Jung$^2$, Friederike Schmid$^3$}
\affiliation{%
$^1$Technische Universit\"{a}t Berlin, Institut f\"{u}r Theoretische Physik, Hardenbergstr. 36, 10623 Berlin, Germany.}%
\affiliation{%
$^2$ Laboratoire Interdisciplinaire de Physique, Université Grenoble Alpes, 38402 Saint-Martin-d'Hères, France}%
\affiliation{%
$^3$Institut f\"{u}r Physik, Johannes Gutenberg-Universit\"{a}t Mainz, 55099 Mainz, Germany.}%

\begin{abstract}
\noindent
Colloidal probes immersed in an active bath have been found to behave like active particles themselves. Here, we use coarse-grained simulations to investigate the mechanisms behind this behavior.
We find that the active motion of the colloid cannot be simply attributed to the convective motion in the bath. Instead, the boundary of the probe contributes significantly to these adopted dynamics by causing active bath particles to spontaneously accumulate at the probe. This gathering of active bath particles then pushes the probe, thus promoting its emergent active-particle-like behavior. 
Furthermore, we find that the dynamic properties of the probe depend on its size in a non-monotonic way, which further highlights the non-trivial interplay between probe and bath.
\end{abstract}

\maketitle

\section{Introduction}
\label{sec:intro}

Many living systems depend on the interplay between active and passive constituents. Important examples include biomixing of fluids and nutrition by microorganisms \cite{Pedley1992activesusp,Kurtuldu2011biomixing} as well as enhanced transport of mesoscopic particles induced by bacteria and motor proteins \cite{Caspi2000enhanced,wu,Mikhailov2015enhanced}. To better understand this interplay between passive and active matter, a vast amount of research has been devoted to understanding the dynamics of a passive probe immersed in an active bath both experimentally~\cite{Leptos_2009_Exp,Ortlieb_2019_Exp,miño_dunstan_rousselet_clément_soto_2013_Exp,wu}, theoretically~\cite{Kanazawa_2020_Theo,Tripathi_2022_Theo,feng2021effective_theo,steff,Ash,Brady,Brady_curved,Voigtmann,Baek_2018,Granek_2020}, and numerically~\cite{ME,Angelani_2009}. A number of these studies have found that a passive probe immersed in an active bath adopts many attributes of active particles themselves \cite{ME,steff,Callegari,Volpe,wu,Voigtmann}, therefore indicating universal emergent active behaviour. Yet it remains unclear which microscopic mechanism leads the probe to acquire these attributes of an active particle. 

The persistent motion of active particles can exert a force perpendicular to the boundary, which accumulates over time and space to generate a `swim pressure', which represents the pressure exerted by the boundary in order to contain the active particles~\cite{Swim_pressure,Silke_2014,Joakim_2015}. For boundaries associated with free assymetric bodies, immersed in the active bath, this swim pressure has been shown to lead to directed motion~\cite{Ash,Brady,Brady_curved,Knežević_2020,Baek_2018,Granek_2020,Angelani_2009,Mallory_2014}. The situation is more complex for symmetric bodies, such as spherical colloids, which will not exhibit persistent directed motion.  However, these symmetric bodies can still experience spontaneous accumulation of active particles at their surfaces which could lead to emergent active propulsion.

Self-propelled particles tend to accumulate at boundaries~\cite{confined_marchetti,Berke_2008,Li_2009,Li_2011,ezhilan_saintillan_2015,Stark_DetentionTimes,Duzgun_2018,flagella_wall}. This accumulation is in part due to complex swimming dynamics of the active particles (e.g. cilia- and flagella-boundary interactions~\cite{flagella_wall} or hydrodynamic interactions~\cite{Berke_2008,Stark_DetentionTimes}) as well as the geometric constraints of the boundary~\cite{Li_2009,Li_2011}. However, the most simple mechanism for this accumulation is the finite persistence time of active particles, which causes them to maintain their orientation for a finite time even after encountering an obstacle. Importantly, this persistent time is much longer than typical inertial time scales, thus rendering the collisions between active particles and the boundary or colloids inelastic and allowing for accumulation of particles. In some sense, the memory of past motion in the active bath is thus transmitted to the passive probe. 



In the present paper, we unravel how the combination of these properties, the accumulation of active particles at boundaries and the directed active motion due to finite persistent time, can lead to spontaneous symmetry breaking and thus to an emergent active behaviour of symmetric passive probes. To determine this mechanism, we consider both impermable and permable spherical passive probes with differnt radii, immersed in a bath of active Langevin particles (ALPs)~\cite{ABP_ALP}. We analyze the inhomogeneous density and orientation of the bath particles in vicinity of the probe and study emergent properties of the passive particle such as its thermal velocity and velocity correlations. Furthermore, we analyze how the memory kernel of the probe changes qualitatively due to the influence of the active bath in driving the colloid out of equilibrium.

 We begin in Section~\ref{sec:alp_like} by examining the active-particle-like behaviors for probes of different sizes. We proceed in Section~\ref{sec:active_fluid} to investigate the properties of the active fluid. In particular, we examine its convective properties in the absence of an immersed probe (Section~\ref{sec:bubb}) and its characteristics in the direct vicinity of an immersed probe (Section~\ref{sec:sph_harm}). In these sections, we also investigate the influence of probe size on these properties. We summarize and conclude in Section~\ref{sec:conc_bi}.

\section{Model and Simulation Details}
\label{sec:syssim}

We consider a three dimensional system of a passive probe immersed in
a bath of active Langevin particles (ALPs) of mass $m$ and radius $R$,
which propel themselves with a constant force $F_0$ subject to
rotational diffusion with a diffusion constant
$D_R$~\cite{ABP_ALP,ALPs_Takatori,
Inertial_delay,Hidden_entropy_Marchetti,Enhanced_diff_Marchetti,MITD,Gompper_local_stress,Loewen_TD_inertia,ME,shea2023force}.  The ALPs are coupled to a thermal bath with temperature $\kbT$ via a Markovian, Langevin thermostat, and they interact with each other and with the immersed probe by repulsive hard core interactions of the Weeks-Chandler-Anderson (WCA) type
\cite{WCA}. 

Specifically, we follow Refs.~\cite{ALPs_Takatori,
Hidden_entropy_Marchetti,ME,shea2023force} and set the rotational inertia of ALPs to
zero for simplicity.  The equations of motion for an ALP, $n$, in the
bath, are thus given by
\begin{equation}
\label{eq:alp_int}
\begin{split}
m \dot{\mathbf{v}}_n(t)&=F_0\mathbf{e}_n(t)-\gamma \mathbf{v}_n(t)+\boldsymbol{\xi}_n(t)\\
&-\nabla U_\mathrm{WCA}(\mathbf{r}_n-\mathbf{R})-\sum_{n\neq m} \nabla U_\mathrm{WCA}(\mathbf{r}_n-\mathbf{r}_m),
\end{split}
\end{equation}
where $F_0$ is the propulsion force of the active particle,
$\mathbf{e}(t)$ is the orientation of the active particle, and
$\gamma=6\pi\eta R$ is the damping constant for an ALP radius $R$ in a
thermal bath with viscosity $\eta$. The term $\boldsymbol{\xi}_n(t)$
represents a stochastic force that mimics implicit collisions of the
ALPs with thermal bath particles. These collisions are modeled as
Gaussian white noise with mean zero and variance given by
a fluctuation-dissipation relation
\begin{equation}
\label{eq:trans_diff}
\langle \xi_i(t)\xi_j(t') \rangle = 2 \gamma \kbT \delta_{ij} \delta (t-t').
\end{equation}
The resulting translational diffusion coefficient of isolated ALPs
is given by $D_T=\kbT/\gamma$. 
Finally, the terms $-\nabla U_\mathrm{WCA}(\mathbf{r}_n-\mathbf{R})$
and $-\sum_{n\neq m} \nabla U_\mathrm{WCA}(\mathbf{r}_n-\mathbf{r}_m)$
in Eq.\ \eqref{eq:alp_int} describe the WCA interactions with the probe
and with all other ALPs, respectively. 

The time evolution of the orientation of the ALP, $\mathbf{e}(t)$, is
governed by rotational diffusion, 
\begin{equation}
\label{eq:rot}
\dot{\mathbf{e}}(t)=\mathbf{N}(t) \times \mathbf{e}(t),
\end{equation}
where $\mathbf{N}(t)$ is again Gaussian white noise with a mean of zero 
and a variance (another fluctuation-dissipation relation)
\begin{equation}
\label{eq:rot_diff}
\langle N_{\alpha}(t)N_{\beta}(t') \rangle = 2 D_R \delta_{\alpha \beta} \delta (t-t').
\end{equation}
Here $D_R$ is the rotational diffusion constant, which is given by
$D_R=3D_T/4R^2$ for a particle of radius $R$. The immersed probe only experiences forces from interactions with the surrounding ALPs. Unlike the ALPs, it is not coupled to the
thermal bath.

All simulations are performed using LAMMPS \cite{lammps}. The length, energy,
and mass scales in the simulation system are defined by the Lennard-Jones (LJ)
diameter $\sigma$, energy $\epsilon$, and ALP mass $m$, respectively,
which defines the LJ time scale $\tlj = \sigma
\sqrt{m/\epsilon}$. We use truncated and shifted LJ potentials with
the energy scale $\epsilon$ which are cut off at
$r_{\mathrm{c}}=2^{\frac{1}{6}}\sigma$, resulting in purely repulsive
WCA interactions as described above. The cubic simulation box has
periodic boundary conditions in all three dimensions and a side length
based on the desired density of the bath. 

In Ref.~\cite{ME}, we have studied a probe of mass $M=100m$ and radius $R_p=3.0\sigma$. We now maintain the mass to volume ratio of the probe from this study --- $M/\mathcal{V}=25/(9\pi) m\sigma^{-3}$ --- but change its radius so that the probe has the parameters shown in Table~\ref{tab:mvratio}.

\begin{table}[h!]
  \centering
  \renewcommand\thetable{\MakeUppercase{\romannumeral 3}.1}
  \begin{tabular}{c | c} 
  $R_p$ & $M$ \\
  \hline
  $0.5\sigma$  & $0.46296m$  \\
  \hline
  $1\sigma$  & $3.7037m$  \\
  \hline
  $2\sigma$  & $29.623m$  \\
  \hline
  $3\sigma$  & $100.00m$  \\
  \hline
  $4\sigma$  & $237.04m$  \\
  \end{tabular}
\caption{List of probe radii ($R_p$) investigated as well as their corresponding masses ($M$), which maintain a constant probe mass to volume ratio of $M/\mathcal{V}=25/(9\pi) m\sigma^{-3}$.}
\label{tab:mvratio}
\end{table}

The body of the probe is
constructed so that its surface is smooth, resulting in full slip
boundary conditions for the LJ fluid. Thus, the bath has no influence
on the rotational motion of the probe. The active bath consists of ALPs
with a mass of $m_{\mathrm{ALP}}=1m$ and a radius of $R=0.5\sigma$.
The number of ALPs in the bath is determined by the desired
density of the bath. The parameters of the thermal bath are chosen
such that $\kbT = 1 \; \epsilon$ and 
$\eta = 1 \; m/(\sigma \tlj)$, resulting in
$\gamma = 3 \pi m/\tlj$, $D_T = (3 \pi)^{-1} \sigma^2/\tlj$,
and $D_R = \pi^{-1} /\tlj$.
The driving force $F_0$ is
varied in the range up to $F_0 \le 50 \: m \sigma^2/t_{lJ}$.

In the following, for consistency with previous work \cite{ME,shea2023force}, we will often also use dimensionless quantities $\htt = t \: \gamma/m$, $\hv = \mathbf{v} \: \sqrt{{m}/{\kbT}}$, $\hf = F_0 \frac{1}{\gamma} \sqrt{{m}/{\kbT}}$, and $\hdr = D_R \: m/\gamma$. Dimensionless quantities will denoted with a hat. Density values and distances are given in LJ units. In these cases, the LJ units are explicitly written with the value.

\section{Active-particle-like behavior for probes of different sizes}
\label{sec:alp_like}

In earlier work~\cite{ME}, we found that a probe
immersed in an active bath exhibits active-particle-like characteristics.
To set the stage for the subsequent analysis of boundary interactions,
we will now begin with characterizing this active-particle-like behavior in detail
for probes of different size.


\subsection{Kinetic temperature}
\label{sec:kin_temp_mvratio}
We first investigate the difference between the kinetic temperature of the probes in Table~\ref{tab:mvratio} and that of the bath ALPs ($\Delta\kbT\equiv\kbT_\mathrm{Probe}-\kbT_\mathrm{ALP\ Bath}$). In Ref.~\cite{ME}, we found that a probe of radius $R_p=3.0\sigma$ exhibits a higher kinetic temperature than that of the bath ALPs. This is a distinct non-equilibrium signature: in equilibrium, we would expect the probe and bath to thermalize to the same kinetic temperature.
Given that our system is out of equilibrium and that we have already seen that the probe kinetic temperature does not necessarily equilibrate to that of the ALP bath, we now analyze how probe size affects kinetic temperature.

Here we define the kinetic temperature of both the probe and a bath ALP as: $k_\mathrm{B}T_\mathrm{eff}=\frac{m}{d}\langle\mathbf{v}^2\rangle$, where $m$ is the particle mass, $d$ is the number of dimensions, and $\langle\mathbf{v}^2\rangle$ is the particle's mean squared velocity~\cite{ME}. We confirm in Appendix~\ref{sec:app_veldist} that, for probes of all different sizes, the probe velocity distribution remains Gaussian in an active bath.


\begin{figure}
  \centering
  \includegraphics[width=\linewidth]{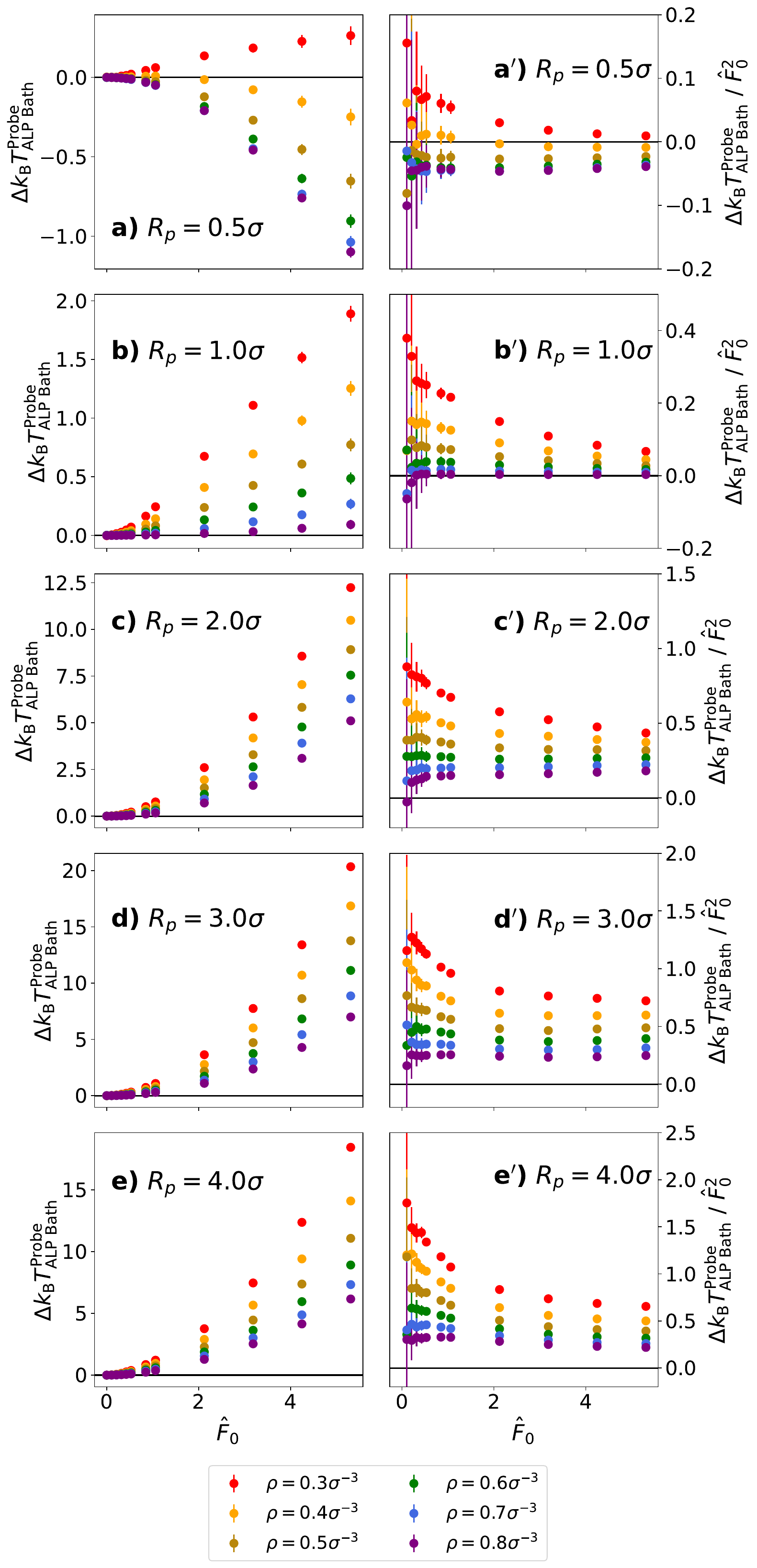}  
\caption{
Difference between the probe kinetic temperature and that of a bath ALP ($\Delta\kbT^{\mathrm{Probe}}_{\mathrm{ALP\ Bath}}=\kbT_\mathrm{Probe}-\kbT_\mathrm{ALP\ bath}$) plotted as a function of the ALP active force ($\hf$) for immersed probes with different radii: \textbf{a}/\textbf{a$^{\prime}$)} $R_p=0.5\sigma$, \textbf{b}/\textbf{b$^{\prime}$)} $R_p=1.0\sigma$, \textbf{c}/\textbf{c$^{\prime}$)} $R_p=2.0\sigma$, \textbf{d}/\textbf{d$^{\prime}$)} $R_p=3.0\sigma$, and \textbf{e}/\textbf{e$^{\prime}$)} $R_p=4.0\sigma$. \textbf{a}-\textbf{d)} The left column shows the unscaled difference, whereas \textbf{a$^{\prime}$}-\textbf{d$^{\prime}$)} the right column shows the difference scaled by $\hf^2$. Different colors show different bath densities ($\rho$). We show a solid black line at $0$.}
\label{fig:temps_mvratio}
\end{figure}

As expected in equilibrium, we find that $\Delta\kbT=0$ for a probe immersed in a passive bath, regardless of bath density and probe size (see Fig.~\ref{fig:temps_mvratio}). However, once the bath becomes active, the
difference becomes nonzero and depends on both the non-equilibrium driving force and the particle radius. In particular, the behavior of $\Delta\kbT$ as a function of $\hf$ for a probe of $R_p=0.5\sigma$ differs significantly from its behavior for larger probes (compare Fig.~\ref{fig:temps_mvratio}a) with Figs.~\ref{fig:temps_mvratio}b-e)). We infer that this different behavior stems from the fact that, in the $R_p=0.5\sigma$ case, the probe radius is the same as that of the ALPs themselves. Therefore, the probe does not pose a significant obstacle and can be shoved relatively easily by the bath ALPs. This is not the case for larger probes, leading to the different qualitative behavior in Fig.~\ref{fig:temps_mvratio}a). 

For all probes with radii $R_p\geq1.0\sigma$, $\Delta\kbT\geq0$ for baths of all densities and activities $\hf>0$, meaning that the probe is `hotter' than the ALP bath (see Figs.~\ref{fig:temps_mvratio}b-e)). Furthermore, in looking at Figs.~\ref{fig:temps_mvratio}b$^{\prime}$-e$^{\prime}$) where we graph $\Delta\kbT/\hf^2$ as a function of $\hf$, we can see that $\Delta\kbT$ scales approximately quadratically with $\hf$ for probes of all radii $R_p\geq1.0\sigma$.
This quadratic scaling of the kinetic temperature difference is the same as was shown for an isolated ALP in Ref.~\cite{ME}. On the other hand, the kinetic temperature difference decreases with increasing bath density, and this again reflects the behavior of the kinetic temperature of the ALP bath itself as a function of density, see Ref.~\cite{ME}: The higher the bath density, the more often the ALP particles undergo random collisions, which effectively slows them down (a bath ALP particle is slower than an isolated ALP). A similar randomization effect seems to govern the interactions with the immersed probe. We will discuss this further in section \ref{sec:active_fluid}.   

\begin{figure}
  \centering
  \includegraphics[width=\linewidth]{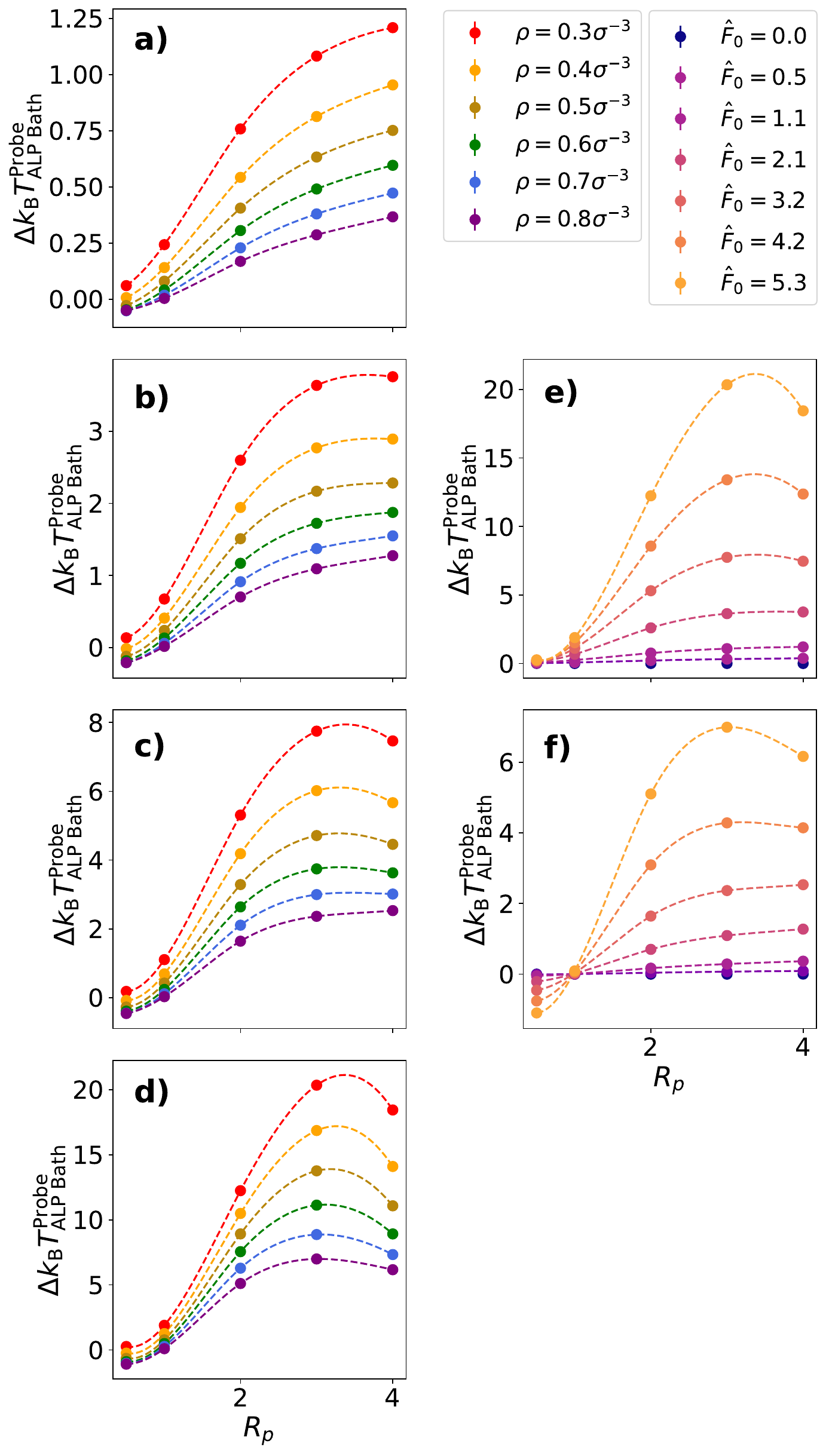}
\caption{Difference between the probe kinetic temperature and that of the ALP bath ($\Delta\kbT^\mathrm{Probe}_\mathrm{ALP\ bath}$) plotted as a function of the probe radius ($R_p$). In the left column, each plot shows ALP baths of different densities ($\rho$, each a different color) for a specific activity ($\hf$): \textbf{a)} $\hat{F}_0=1.1$, \textbf{b)} $\hat{F}_0=2.1$, \textbf{c)} $\hat{F}_0=3.2$, and \textbf{d)} $\hat{F}_0=5.3$. In the right column, each plot shows ALP baths of different activities ($\hf$, each a different color) for a specific bath density ($\rho$): \textbf{e)} $\rho=0.3\sigma^{-3}$ and \textbf{f)} $\rho=0.8\sigma^{-3}$.}
\label{fig:temps_rad}
\end{figure}

%

\begin{figure}
  \centering
  \includegraphics[width=\linewidth]{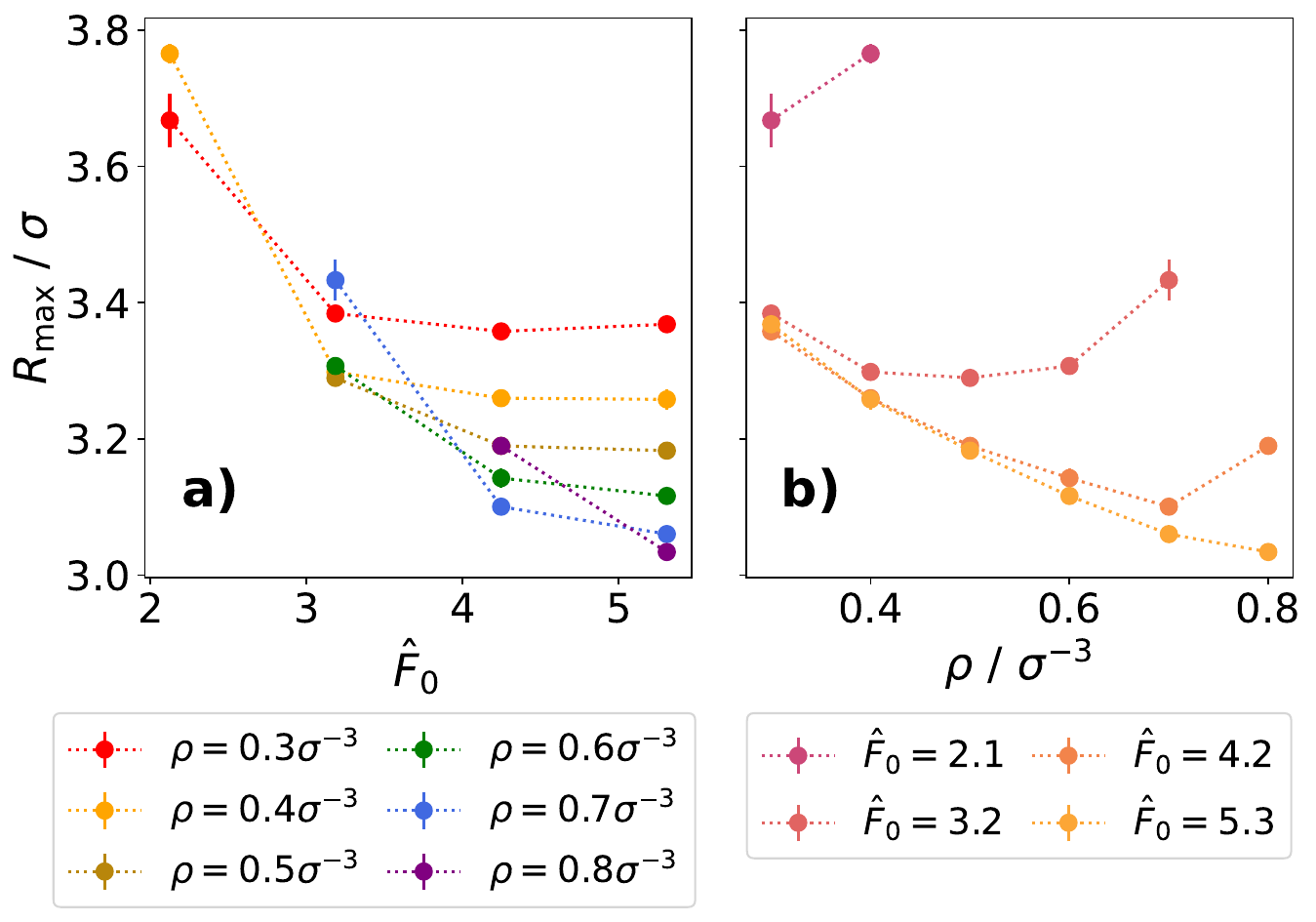}  
\caption{Radius of the probe ($R_p$) at which $\Delta\kbT$ reaches a maximum as determined by spline interpolation between the data points in Fig.~\ref{fig:temps_rad}. \textbf{a)} Maximum as a function of the active force $\hf$ where different colors represent different densities ($\rho$). \textbf{b)} Maximum as a function of the bath density $\rho$ where different colors represent different activities ($\hf$). Densities in the legend are expressed in LJ units of $\sigma^{-3}$.}
\label{fig:spline_max}
\end{figure}

In Fig.~\ref{fig:temps_rad}, we replot the same data as in Fig.~\ref{fig:temps_mvratio}, but now showing $\Delta\kbT^\mathrm{Probe}_\mathrm{ALP\ bath}$ as a function of particle radius $R_p$, and here, a surprise occurs: For sufficiently high bath activities, the curves are non-monotonic and exhibit a maximum. This is observed for
ALP baths of all densities; 
however, the active force at which the non-monotonicity begins depends on the density. We also see in Fig.~\ref{fig:temps_rad} that the maximum value of $\Delta\kbT$ always occurs around $R_p=3\sigma$, independent of bath density and activity. To better localize
the maximum, we fit $\Delta\kbT$ as a function of $\hf$ with a spline interpolation (shown as dotted lines in Fig.~\ref{fig:temps_rad}) and calculate the maximum of this interpolation ($R_\mathrm{max}$), if one exists. The results are shown in Fig.~\ref{fig:spline_max}. In creating Fig.~\ref{fig:spline_max}, we only consider activities $\hf>0.2$ because of larger statistical uncertainties in the value of $\Delta\kbT$ for very low activities.

Fig.~\ref{fig:spline_max} confirms that the non-monotonic behavior (i.e. the presence of a maximum $R_{\textrm{max}}$) occurs at all densities, but only at higher active forces. For lower active forces, we do not see non-monotonic behavior; however, we cannot exclude the possibility that a maximum may exist beyond our largest studied radius, $R_p=4.0\sigma$. Fig.~\ref{fig:spline_max}a) shows that, the higher the density of the bath, the higher the activity necessary for non-monotonic behavior to occur. We also see that, based on our interpolation, that
$R_\mathrm{max}$ slightly depends  on both the activity and density of the bath. According to Fig.~\ref{fig:spline_max}a), $R_\mathrm{max}$ initially decreases with increasing bath activity and then plateaus. The plateau value of $R_\mathrm{max}$ decreases with increasing density. 

The observation of this maximum in the kinetic temperature is unexpected, and the explanation is not obvious. For example, one might suspect that $R_\textrm{max}$ is
related to a typical distance covered by an ALP at the surface of the probe 
before the latter reorients. However, in this case it should be correlated with the persistence length of the ALPs, which increases strongly with increasing active force 
(see Appendix~\ref{sec:app_pl}), whereas $R_\textrm{max}$ remains roughly constant. 
Alternatively, one might suspect  $R_\textrm{max}$ to be connected with the correlation length of collective velocity alignment in the ALP bath. However, as we show in 
Appendix \ref{sec:app_corr}, the range of these correlations is much smaller than
$R_\textrm{max}$. In section \ref{sec:active_fluid}, we will reconsider this puzzle
by analysis of the ALP fluid structure at the probe boundaries.



We conclude this section with the comment that the significant dependence of the probe dynamics on its own boundary provides us with another non-equilibrium signature for a probe immersed in an active bath. In an equilibrium system, we would expect that probes of all sizes would equilibrate to the same kinetic temperature. Therefore, the fact that probes of different sizes achieve different kinetic temperatures in an active bath can be used as a non-equilibrium signature. In Ref.~\cite{ME}, we proposed one non-equilibrium signature of a probe immersed in an active bath which can be determined only from the probe trajectory: the violation of the first fluctuation dissipation theorem. We now propose another non-equilibrium signature of this system which can be determined from the trajectories of differently sized probes: different kinetic temperatures for probes of different sizes in the same active bath. However, this new non-equilibrium signature requires the simultaneous study of two probes and cannot be identified based on one probe trajectory alone.

\subsection{Velocity autocorrelation function and memory kernel}
\label{sec:vacf_mk_mvratio}

To uncover further active-particle-like properties exhibited by probes of different sizes, we now investigate their dynamic properties. We first focus on the velocity autocorrelation function (VACF). It has been shown that the VACF of both an isolated ALP and an individual constituent of an ALP fluid decay exponentially with $2D_R$ in the limit $t\to\infty$~\cite{ME}. 

For large probes immersed in low activity baths, the values of the VACF in the long time limit are too close to zero to allow for a reliable estimation of decay rates.
For small probes immersed in low activity baths and large probes immersed in high activity baths, the VACF decays exponentially at a rate which is very similar to that of the bath ALPs. 
However, for the case of small probes immersed in high activity baths the decay rate is clearly not the same as that of the bath ALPs: the decay rate is instead higher. 
We infer that this is due to the fact that the small probes constantly collide with constituents of the active fluid, but do not have sufficient inertia to maintain their current direction. This leads to an effectively lower persistence length of the probe and thereby a faster decay of the VACF. This different behavior of the small probes is not surprising, given the different behavior of the probe kinetic temperature which we already saw in Section~\ref{sec:kin_temp_mvratio}.

Thus, at least in certain cases, the probe acquires VACF behavior reminiscent of that of an ALP in the surrounding bath. We now further investigate the dynamic behavior of the probe by mapping its
movement onto a generalized Langevin equation (GLE)~\cite{Izvekov2013_PO,memory_review,proj_op,wip,Izvekov2013_PO_NEQ,Vroylandt_2022_FDT,netz2023derivation}:
\begin{equation}
\label{eq:gle}
M \dot{\mathbf{V}}(t)=-\int^t_0 \mathrm{d}s \: K(t-s) \mathbf{V}(s)+\bm{\Gamma}(t),
\end{equation}
where $M$ is the mass of the probe, $\mathbf{V}(t)$ is its velocity,
$K(t-s)$ is the memory kernel, and $\bm{\Gamma}(t)$ is the stochastic
force on the probe. In mapping the motion of the probe to the GLE, we
explicitly allow that the effective dynamics of the colloid in the ALP
fluid may be non-Markovian because previous studies have shown that, under certain conditions, the dynamics of a probe immersed in an active bath cannot be assumed to be Markovian~\cite{Voigtmann,abbasi2022non}. The memory kernel $K(t)$ is determined
from the VACF by Volterra inversion, as suggested by the Mori-Zwanzig
projection operator formalism~\cite{Zwanzig_Non,MZ,Zwanzig_Mem,Shin_2010,Carof_2014,memory_review}.
We emphasize that the GLE is a coarse-grained model equation, not the true
dynamical equation of motion for the probe in the explicit active
fluid \cite{ME,shea2023force}. 

\begin{figure}
  \centering
  \includegraphics[width=\linewidth]{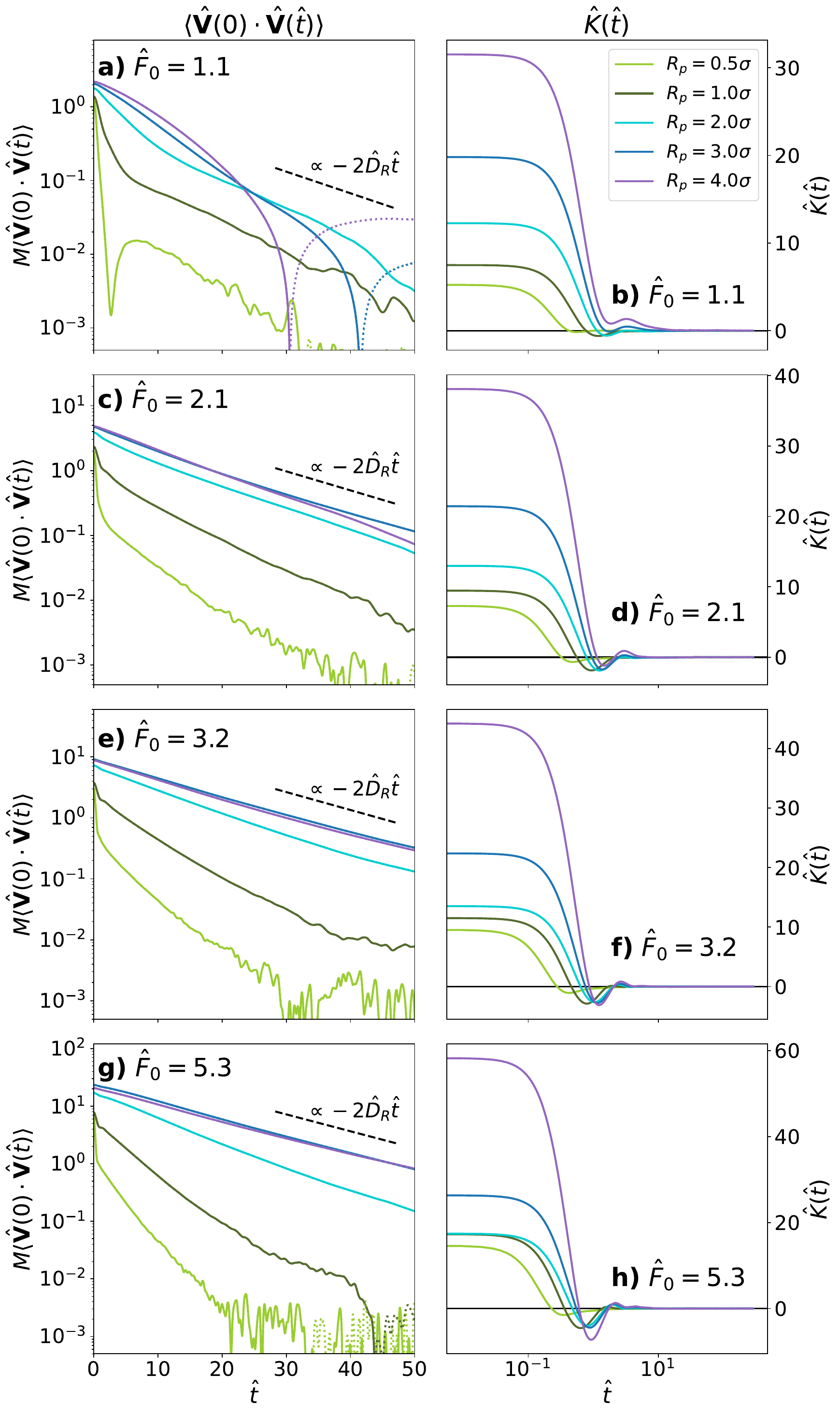}
\caption{VACF and memory kernel. \textbf{a)}, \textbf{c)}, \textbf{e)}, \textbf{g)} Semi-logarithmic plot of the VACF for probes of different radii. Each graph shows a different bath activity. Dotted lines show the absolute value of the VACF. \textbf{b)}, \textbf{d)}, \textbf{f)}, \textbf{h)} Semi-logarithmic plot of the memory kernel for probes of different radii. Each graph shows a different bath activity. The active bath has a density of $\rho=0.4\sigma^{-3}$ for all cases.}
\label{fig:vacf_mem}
\end{figure}

In Fig.\ \ref{fig:vacf_mem}, we see that the shape of the memory kernel
changes qualitatively with increasing activity level in the ALP fluid.
It becomes non-monotonous and negative at intermediate times,
indicating transient positive feedback that promotes superdiffusive
behavior. This is also observed in the memory kernel of isolated ALPs and
consistent with our previous findings
for a probe of $R_p=3.0\sigma$~\cite{ME}. Although differences between memory kernels of probes with various sizes are clearly visible, the behavior remains qualitatively the same with a short-time decay and a pronounced minimum at intermediate times for sufficiently high activities. The active force at which this negative portion of the memory kernel first appears is however probe size dependent, as well as its depth. Interestingly, we see in Fig.~\ref{fig:vacf_mem} that the value of $\hat{K}(0)$ increases monotonically with the radius at all activities. This is in contrast to the initial value of the VACF, which we know -- based on the kinetic temperature -- becomes non-monotonous for large activities.

\section{Characteristics of the active fluid and boundary effects}
\label{sec:active_fluid}


We have now seen that probes of all sizes $R_p\geq1.0\sigma$ exhibit active-particle-like behavior in that they exhibit an enhanced kinetic temperature which scales quadratically with the bath activity and their memory kernels indicate transient positive feedback (this behavior of the memory kernel also occurs for $R_p=0.5\sigma$). In certain cases, the VACF of the probe also adopts the properties of the surrounding ALP particles. Furthermore, we have seen that the diferrence between the kinetic temperature of probe and the ALP bath particles, $\Delta\kbT$, has a non-monotonic relationship with the probe radius $R_p$. We now proceed to investigate the mechanism underlying this active-particle-like behavior as well as the cause of the non-monotonic behavior of $\Delta\kbT$ as a function of the probe radius.

\subsection{Convective properties}
\label{sec:bubb}

\begin{figure}
  \centering
  \includegraphics[width=.7\linewidth]{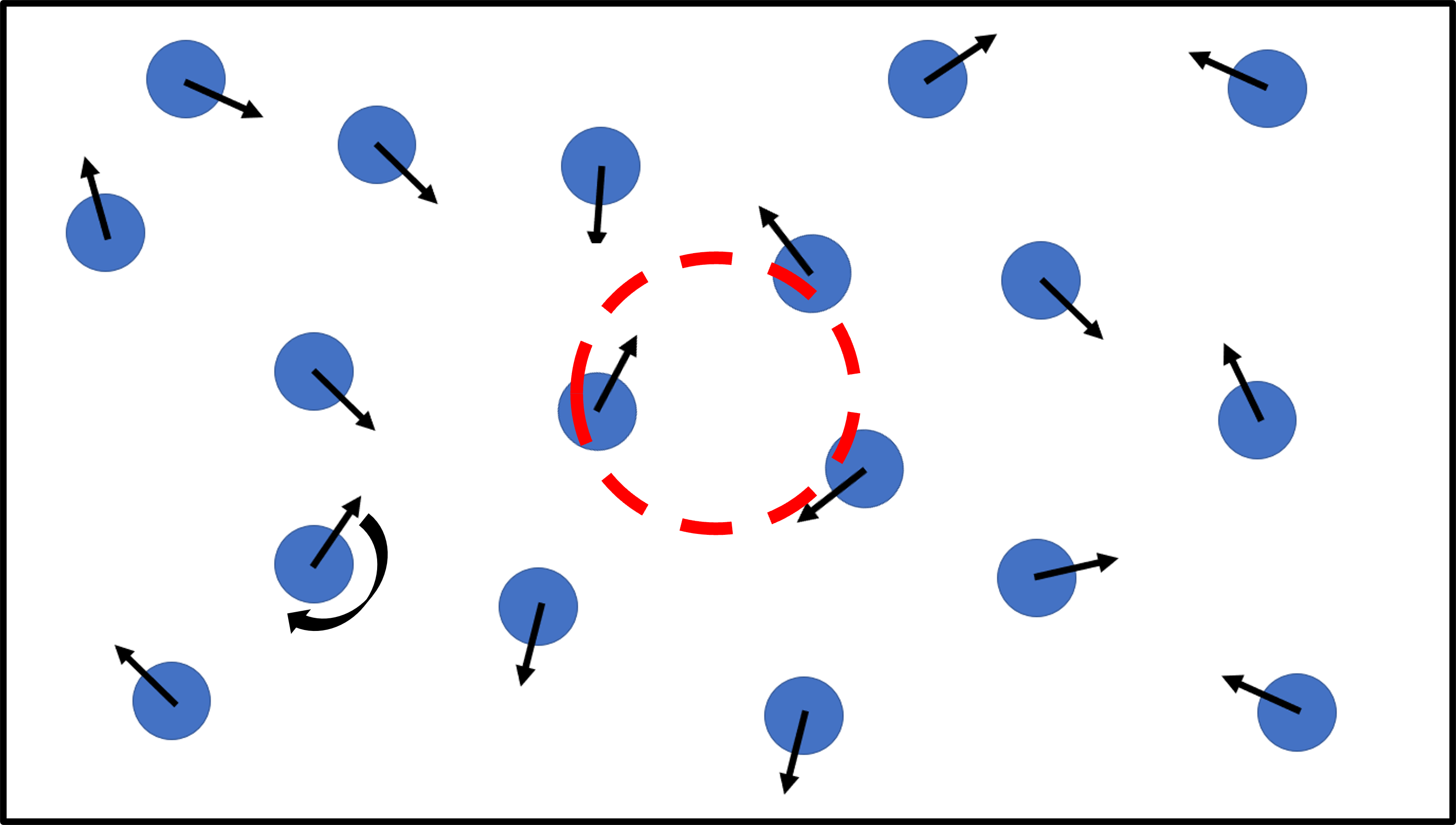}  
\caption{System of a transparent, convective probe (`bubble') immersed in a bath of active Langevin particles.}
\label{fig:bubble_scheme}
\end{figure}

We have seen that all probes $R_p\geq1\sigma$ exhibit an enhanced kinetic temperature, which scales quadratically with the bath activity, similar to an ALP itself. But is this active-particle-like behavior purely due to the convective properties of the active bath, or does the interface between the probe and the active bath contribute to this behavior? In order to answer this question, we again consider a bath of ALPs with mass $m$ and velocity $\mathbf{v}_i$ in three dimensions. As stated in Section~\ref{sec:syssim}, the kinetic temperature of a bath ALP is given by $k_\mathrm{B}T_\mathrm{ALP\ bath}=\frac{m}{d}\langle\mathbf{v}^2\rangle$, where $d$ is the dimensionality of the system; hence,
$\langle\mathbf{v}^2\rangle=\frac{d}{m}k_\mathrm{B}T_\mathrm{ALP\ bath}$. We now consider a transparent, convective probe particle (`bubble') immersed in the bath (see Fig.~\ref{fig:bubble_scheme}). This bubble covers a volume $\mathcal{V}=4/3\pi R_p^3$ and swims with the bath ALPs without otherwise affecting them. Namely, if ALPs $j=1...N$ are contained in the volume $\mathcal{V}$ of the bubble, then its velocity is $\mathbf{V}=\frac{1}{N}\sum^N_{j=1}\mathbf{v}_j$ and its squared velocity is $\langle\mathbf{V}^2\rangle=\frac{1}{N^2}\sum_{jk}\langle\mathbf{v}_j\mathbf{v}_k\rangle$. In the simplest case, if the ALPs are totally uncorrelated, then $\langle\mathbf{V}^2\rangle=\frac{1}{N}\langle\mathbf{v}^2\rangle=\frac{d}{Nm}k_\mathrm{B}T_\mathrm{ALP\ bath}$.

We now assign a hypothetical mass $M$ to the bubble, so that we can
determine its hypothetical kinetic temperature, $k_\mathrm{B}T_\mathrm{Probe}$. For the uncorrelated ALP particles, we obtain $T_\mathrm{Probe}=\frac{M}{Nm}T_\mathrm{ALP\ bath}$. For an ALP bath of density $\rho$, the number of particles contained in volume $\mathcal{V}$ is $N=\rho\mathcal{V}$. Thus, for uncorrelated ALPs:

\begin{equation}
\frac{\rho T_\mathrm{Probe}}{T_\mathrm{ALP\ bath}}=\frac{M}{m\mathcal{V}}.    \label{eq:bubble}
\end{equation}
For comparability purposes, we now consider bubbles with the radii and corresponding hypothetical masses listed in Table~\ref{tab:mvratio}. From Eq.~\eqref{eq:bubble}, given our ALP mass and the probe mass to volume ratio (which is kept constant as $M/\mathcal{V}=25/(9\pi) m\sigma^{-3}$ for probes of all radii), we expect that $T_\mathrm{Probe}>T_\mathrm{ALP\ bath}$ for a bubble in a bath of uncorrelated ALPs with $\rho<1\sigma^{-3}$. This relation is thus qualitatively consistent with the results in Ref.~\cite{ME} and Section~\ref{sec:kin_temp_mvratio} for a hard probe. Therefore, without further investigation, it could be inferred that the hard probe's enhanced kinetic temperature might simply be due to the convective properties of the bath. However, a quantitative inspection shows that the difference $\kbT$ of the probe and ALP bath temperature greatly exceeds that predicted by Eq. (\ref{eq:bubble}.

\begin{figure}
  \centering
  \includegraphics[width=0.95\linewidth]{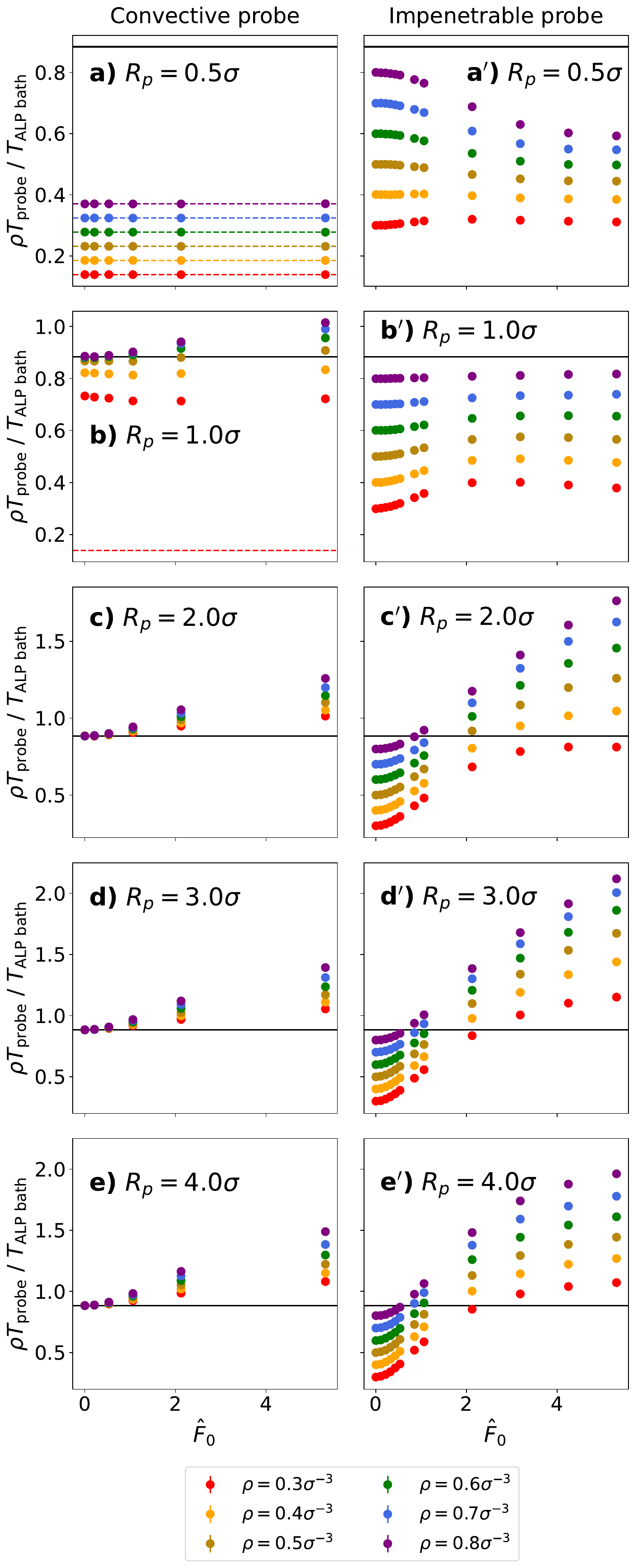}  
\caption{
Ratio of the probe temperature ($T_\mathrm{Probe}$) to that of a bath ALP ($T_\mathrm{ALP\ bath}$) scaled by the bath density ($\rho$) and plotted as a function of the ALP active force ($\hf$) for immersed probes with different radii: \textbf{a}/\textbf{a}$^{\prime}$) $R_p=0.5\sigma$, \textbf{b}/\textbf{b}$^{\prime}$) $R_p=1.0\sigma$,  \textbf{c}/\textbf{c}$^{\prime}$) $R_p=2.0\sigma$, \textbf{d}/\textbf{d}$^{\prime}$) $R_p=3.0\sigma$, and \textbf{e}/\textbf{e}$^{\prime}$) $R_p=4.0\sigma$. \textbf{a}-\textbf{e)} For a transparent, convective probe. \textbf{a}$^{\prime}$-\textbf{e}$^{\prime}$) For an impenetrable probe with a hard boundary. The solid black line in each graph shows the value predicted for a bath of uncorrelated ALPs (Eq.~\eqref{eq:bubble}). The colored, dotted lines in \textbf{a}$^{\prime}$) show $\rho M/m$. We have also included this line for $\rho=0.3\sigma^{-3}$ in \textbf{b}$^{\prime}$) for reference. 
}
\label{fig:temps_bubb}
\end{figure}

On the other hand, our ALPs are not necessarily uncorrelated. To assess the correlations within the ALP bath, we calculate $\Lambda_\mathrm{Bubble}\equiv\rho T_\mathrm{Probe}/T_\mathrm{ALP\ bath}$ for simulations of a transparent, convective probe immersed in an ALP bath. We then compare these values to the theoretical value for a bath of uncorrelated ALP particles (Eq.~\eqref{eq:bubble}) in Figs.~\ref{fig:temps_bubb}a-e).

We first remark that analyzing a bubble of radius $R_p=0.5\sigma$ (Fig.~\ref{fig:temps_bubb}a)) does not provide us with additional insight because, in this case, the volume of the bubble is identical to that of an ALP. Therefore, the bubble simply tracks a singular bath ALP. We thus expect that $T_\mathrm{Probe}/M=T_\mathrm{ALP\ Bath}/m$, so $\Lambda_\mathrm{Bubble}=\rho M/m=\frac{24}{54} \rho$ for all values of activity, which agrees with our results in Fig.~\ref{fig:temps_bubb}a). 

In Fig.~\ref{fig:temps_bubb}b), we see that the behavior of a bubble with radius $R_p=1.0\sigma$ also shows qualitatively different behavior from bubbles with larger radii. We infer that this behavior is due to the small bubble volume. In high density baths, the bubble behavior approaches that of larger bubbles because the bath is sufficiently dense that multiple particles are encompassed within the bubble volume. However, in low density baths, the bubble encompasses, at most times, only one ALP and, consequently, its behavior approaches that of a single ALP.

For bubbles with radii $R_p>1.0\sigma$ in passive and low activity baths, we find that $\Lambda_\mathrm{Bubble}$ calculated from simulation data matches that predicted for a bath of uncorrelated particles (Eq.~\eqref{eq:bubble}) very well, regardless of bath density. This indicates that, in our bath model, for passive and low activity baths, the motion of constituent particles is uncorrelated. However, as the activity of the bath increases, we see that our theoretical model for the bubble underestimates the value of $\Lambda_\mathrm{Bubble}$, indicating that correlations between bath particles emerge with increased activity. These correlations become even more pronounced in higher density baths.

We now assess how the boundary of the probe affects the correlations among bath ALPs by calculating the ratio $\Lambda_\mathrm{Probe}\equiv\rho T_\mathrm{Probe}/T_\mathrm{ALP\ bath}$ for a hard probe (see Figs.~\ref{fig:temps_bubb}a$^{\prime}$-e$^{\prime}$)). We consider hard probes with radius and mass parameters listed in Table~\ref{tab:mvratio}. For a hard probe immersed in a passive bath, the probe and bath will be in thermal equilibrium such that $T_\mathrm{Probe}=T_\mathrm{bath}$; therefore, $\Lambda_\mathrm{Probe}=\rho$, as is shown for passive probes of all sizes immersed in passive baths of all densities in Fig.~\ref{fig:temps_bubb}.


Given that, for bubbles of $R_p=0.5\sigma$ and $R_p=1.0\sigma$, the bubble often only tracks a single ALP particle, comparing $\Lambda_\mathrm{Probe}$ and $\Lambda_\mathrm{Bubble}$ for probes of these sizes does not reveal the effects of the probe boundary on probe dynamics. Therefore, we do not discuss a comparison between $\Lambda_\mathrm{Probe}$ and $\Lambda_\mathrm{Bubble}$ for probes of $R_p=0.5\sigma$ and $R_p=1.0\sigma$. We do, however, note that the behavior of $\Lambda_\mathrm{Probe}$ as a function of $\hf$ for small probes ($R_p\leq1.0\sigma$) qualitatively differs from that which we see for probes of larger radii. We infer that these qualitative differences result from the fact that the probe and the ALPs are on the same length scale.

For all probes with $R_p>1.0\sigma$, $\Lambda_\mathrm{Probe}<\Lambda_\mathrm{Bubble}$ for low bath activities. For high bath activities, $\Lambda_\mathrm{Probe}>\Lambda_\mathrm{Bubble}$. This means that, for low bath activities, the probe boundary \emph{anti-correlates} the bath particles. In the passive case, this anti-correlation is necessary for the probe to come to thermal equilibrium with the bath. For high bath activities, on the other hand, the probe boundary \emph{correlates} the bath particles.

The most important, if perhaps also most general, conclusion to be drawn from comparing $\Lambda_\mathrm{Probe}$ to $\Lambda_\mathrm{Bubble}$ is that --- for a given bath density, bath activity, and probe radius --- they are not the same (i.e. Figs.~\ref{fig:temps_bubb}a$^{\prime}$-e$^{\prime}$) are not equivalent to Figs.~\ref{fig:temps_bubb}a-e)). Therefore, the dynamics of the immersed hard probe cannot solely be attributed to the convective properties of the bath. Rather, the probe boundary strongly affects the correlations within the ALP bath, which then influence the probe behavior.

\subsection{Properties in the probe vicinity}
\label{sec:sph_harm}
Given that the probe boundary plays an important role in both the dynamics of the immersed probe and the correlations in the ALP bath, we now examine the properties of the ALP fluid in the vicinity of the probe. We do this for probes of all sizes listed in Table~\ref{tab:mvratio}. We examine the angle-dependent density distribution $\rho(\mathbf{r})$ of ALPs around the probe in a comoving and corotating frame with origin at the probe center, $\mathbf{R}(t)$, and $z$ axis always aligned in the direction of the instantaneous probe velocity $\mathbf{V}(t)$. This method has already been used to study the properties of an ALP fluid surrounding a probe of radius $R_p=3.0\sigma$ in Ref.~\cite{shea2023force}. 

To quantify the angular dependence, we expand the density distribution $\rho(\mathbf{r})$ 
in spherical harmonics,
\begin{equation}
 \rho(\mathbf{r}) = \sum_{lm} Y_{lm}^*(\mathbf{r}/r) \: \Omega_l^m(r).
\end{equation}
The coefficients $\Omega_l^m(r)$ can be determined from simulations according to 
\begin{equation}
\Omega^m_l(r)=\frac{1}{\mathcal{V}}\sum_{n\in \delta \mathbf{r}}Y_{lm}(\mathbf{r}_n/r_n),
\label{eq:yl1m0}
\end{equation}
where the sum $\sum_{n\in \delta \mathbf{r}}$ runs over all bath particles in a spherical shell $\delta r$ 
around the particle (i.e., their distance from the probe center lies within the interval
$[r-\delta r/2: r + \delta r/2]$), and $\mathcal{V}$ is the volume of the shell, $\mathcal{V}(r)=4/3\pi((r+\delta r/2)^3-(r-\delta r/2)^3)$. Specifically, the radial average of $\rho(\mathbf{r})$ can be obtained from
\begin{equation}
\rho(r)= Y_{00}^* \: \Omega_0^0 = \frac{\sqrt{4\pi}}{\mathcal{V}}\sum_{n\in \delta \mathbf{r}}Y^0_0(\mathbf{r}_n/r_n).
\label{eq:density}
\end{equation} 

\begin{figure*}
  \centering
  \includegraphics[width=\linewidth]{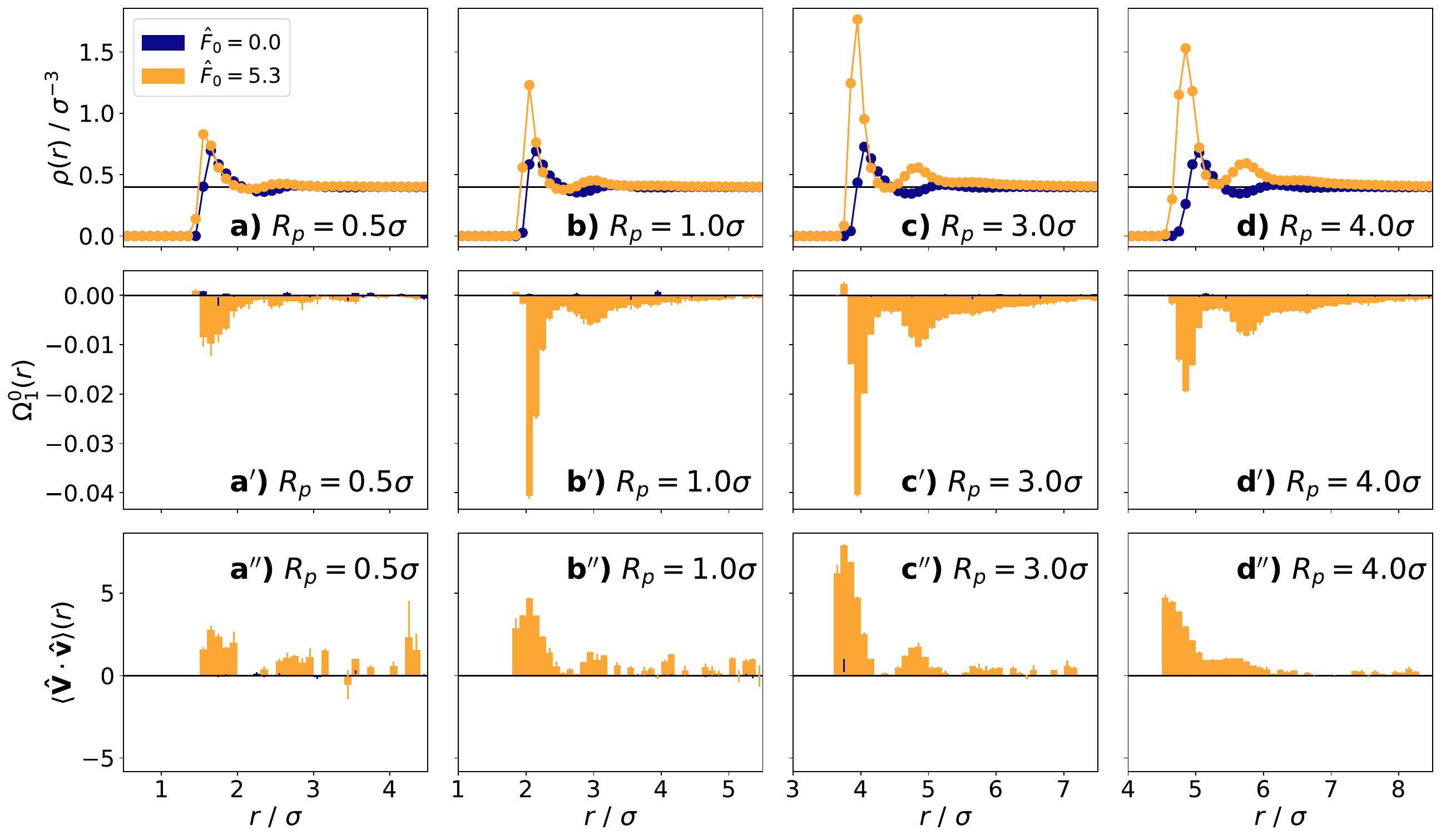}  
\caption{\textbf{a}-\textbf{d}) Density of the bath ($\rho(r)$) in LJ units of $\sigma^{-3}$, \textbf{a}$^{\prime}$-\textbf{d}$^{\prime}$) dipole moment of the bath ($\Omega_1^0(r)$, see Eq.~\eqref{eq:yl1m0}), and \textbf{a}$^{\prime\prime}$-\textbf{d}$^{\prime\prime}$) alignment of the ALP velocities with that of the probe velocity ($\langle\mathbf{V}\cdot\mathbf{v}\rangle(r)$, see Eq.~\ref{eq:velcorr}) as a function of distance from the center of the immersed probe ($r$) for baths of average density $\rho_0=0.4\sigma^{-3}$. Each column shows a probe of a different radius: \textbf{a}/\textbf{a}$^{\prime}$/\textbf{a}$^{\prime\prime}$) $R_p=0.5\sigma$, \textbf{b}/\textbf{b}$^{\prime}$/\textbf{b}$^{\prime\prime}$) $R_p=1.0\sigma$,  \textbf{c}/\textbf{c}$^{\prime}$/\textbf{c}$^{\prime\prime}$) $R_p=3.0\sigma$, and \textbf{d}/\textbf{d}$^{\prime}$/\textbf{d}$^{\prime\prime}$) $R_p=4.0\sigma$. For ease of visualization, in \textbf{a}-\textbf{d}) we show a black line at the global density and in \textbf{a}$^{\prime}$-\textbf{d}$^{\prime}$) and \textbf{a}$^{\prime\prime}$-\textbf{d}$^{\prime\prime}$) we show a black line at $0$.
}
\label{fig:yl01m00}
\end{figure*}

In Ref.~\cite{Brady}, it was analytically found that, for a sphere immersed in a dilute 3D system of ABPs, the ABP concentration decays exponentially to the bulk density. 
We now examine the density profile surrounding a spherical probe immersed in a dense 3D system of ALPs through simulation data. We can see in Figs.~\ref{fig:yl01m00}a-e) that, even in a bath of density $\rho=0.4\sigma^{-3}$, our bath is sufficiently dense to show oscillations before approaching the bulk density. 

In comparing the density curves for a passive and an active bath in Figs.~\ref{fig:yl01m00}a-e), we see that the first peak of the density curve is higher in an active bath for probes of all radii. This means that adding activity to the bath leads to a higher density of bath particles in the vicinity of the probe. Interestingly, we see that the magnitude of increase in the peak density is non-monotonic with the probe radius. Similarly to the behavior of $\Delta\kbT$, this difference reaches a maximum for a probe of $R_p=3.0\sigma$ and then decreases for a probe of $R_p=4.0\sigma$.

We also see that the initial peak of the active bath density profile is shifted slightly closer to the probe center in comparison with that of the passive bath, again independent of probe radius. Furthermore, the spacing between the first and second shells of the active bath particles are closer together. These closer shells indicate that ALPs are able to move closer to the probe, and each other, than passive bath particles. We infer that this ability stems from the higher kinetic energy of ALPs due to their active force, which allows them to overcome more of the repulsive potential from WCA interactions with the probe.

For a passive bath, the higher order spherical harmonics of the density distribution of the bath are all zero. On the other hand, from Figs.~\ref{fig:yl01m00}a$^{\prime}$)-d$^{\prime}$), we see that the ALP bath acquires a negative dipole moment surrounding the probe, which is sustained to large values of $r$. This negative dipole moment indicates that ALPs collect behind the probe relative to its instantaneous velocity, even at large distances from the probe. The phenomenon of active particles gathering behind an immersed passive probe has previously been seen for a probe dragged through an ABP bath in Ref.~\cite{Stark_Milos}, where it is framed as a difference between the forces behind and in front of the probe.
In our systems with  spherically symmetric probes, we find that the spatial distribution of ALP particles can be fully characterized by the $\rho(r)$ and $\Omega_1^0(r)$. Higher order spherical harmonics of the density distribution did not exhibit structure within the error, neither in the active, nor in the passive bath.

We note here that, as seen in the density profile, the magnitude of the dipole moment peak has a non-monotonic relationship with the probe radius.
The small values of $\Omega^0_1(r)$ for a probe of radius $R_p=0.5\sigma$ demonstrate that the accumulation mechanism is not significant for a probe of this size. The failure of this mechanism can be explained intuitively: If the probe has the same size (in volume) as the ALPs themselves, 
significant accumulation is not possible. This lack of accumulation helps to explain why the behavior of $\Delta\kbT$ as a function of $\hf$ for a probe of $R_p=0.5\sigma$ differs significantly from probes with larger radii (see Fig.~\ref{fig:temps_mvratio}).
Large probes with $R_p=4.0\sigma$ also have smaller values of $\Omega^0_1(r)$ when compared with probes of intermediate sizes. This can be understood by considering the limit that the size (mass) of the probe becomes infinite. In this case, there will also be no net dipole moment because the inertia of the probe will dominate and the particle will not move. Therefore, combining this knowledge of the small and large probe limits, we conclude that there must be a maximum in between, i.e. the magnitude of the dipole moment must exhibit non-monotonic behavior.

We infer that the velocities of the ALPs collected behind the probe become correlated and effectively \emph{push} the probe.
To assess this hypothesis, we calculate the alignment of ALP velocities with that of the probe as: 

\begin{equation}
\langle \mathbf{V}\cdot\mathbf{v}\rangle(r)=\frac{1}{\rho(r)}\sum_{i\in \delta \mathbf{r}}\mathbf{V}\cdot\mathbf{v}_i.
\label{eq:velcorr}
\end{equation}

Figs.~\ref{fig:yl01m00}\textbf{a}$^{\prime\prime}$-\textbf{d}$^{\prime\prime}$), confirm that the probe velocity is in fact aligned with the ALPs in its vicinity, supporting our hypothesis of this pushing mechanism. We see in Figs.~\ref{fig:yl01m00}\textbf{a}$^{\prime\prime}$-\textbf{d}$^{\prime\prime}$) that the height of the maximum peak in $\langle \mathbf{V}\cdot\mathbf{v}\rangle(r)$ has a non-monotonic relationship to the probe radius, much like $\Delta\kbT$ and the maximum of the density peak. The maximum peak of $\langle\mathbf{V}\cdot\mathbf{v}\rangle(r)$ occurs, once again, at approximately $R_p=3\sigma$.

\begin{figure}
  \centering
  \includegraphics[width=\linewidth]{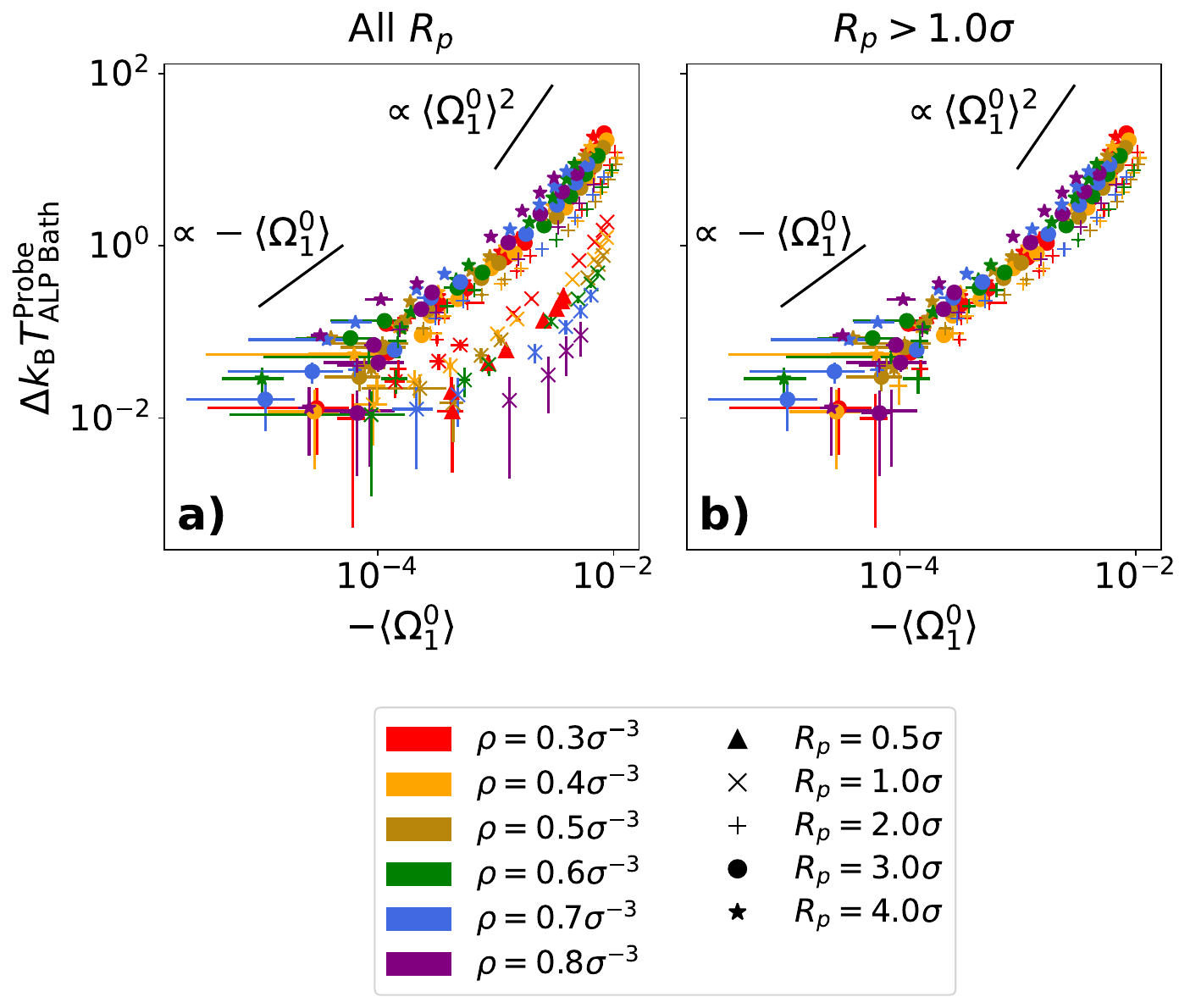}  
\caption{Kinetic temperature difference between the immersed probe and the ALP bath as a function of the average dipole moment, $\langle \Omega^0_1\rangle$, for \textbf{a)} all radii and \textbf{b)} for radii $R_p>1.0\sigma$}. Data for all simulated bath densities ($\rho$) and activities ($\hf$), as well as for all probe radii ($R_p$), are shown. Different probe radii a distinguished by different symbols and different bath densities are distinguished by different colors. Bath activities are not distinguished.
\label{fig:dip_temp}
\end{figure}

We furthermore infer that the pushing by the ALPs promotes the enhanced kinetic temperature of the probe and induces its active-particle-like behavior (i.e. its VACF and memory kernel behavior). Indeed, we can see in Fig.~\ref{fig:dip_temp} that the kinetic temperature difference between the immersed probe and the ALP bath has a strong positive correlation with the average dipole moment, $\langle\Omega^0_1\rangle$ -- for small values approximately linearly and for large values approximately quadratically. Here $\langle\Omega^0_1\rangle$ is defined as

\begin{equation}
    \begin{split}
    \langle\Omega^0_1\rangle &=  \int_{\mathcal{V}_1} \mathrm{d}\mathcal{V}~\Omega_1^0(r)/(4\pi r^2) \\        
    \end{split}
\end{equation}
where $\mathcal{V}_1$ is the volume corresponding to the first peak in the dipole moment. The integral on the right hand side of the equation was found to be positive in only very few cases for low values of $\hf$.


Fig.~\ref{fig:dip_temp}a shows a double logarithmic plot of the
kinetic energy difference between probe particle and bath, $\Delta k_B T^{\text{Probe}}_{\text{ALP bath}}$, versus  $(- \langle\Omega^0_1\rangle)$. Points where $\langle \Omega^0_1\rangle$
and/or $\Delta \kbT$
are zero within the error are omitted.
 We note a strong positive correlation
between $\Delta k_B T^{\text{Probe}}_{\text{ALP bath}}$ and $- \langle\Omega^0_1\rangle^2$ for probes of radii $R_p>1.0\sigma$. For  $R_p=0.5\sigma$, most values of
$\langle \Omega^0_1\rangle$ are close to zero and hence not shown.
For probes with $R_p=1.0\sigma$, we see a cluster of points in the lower right corner of the graph. The location of this cluster indicates that, in spite of high values of $\langle\Omega_1^0\rangle$, $\kbT$ remains low. We infer that this weaker coupling between $\langle\Omega_1^0\rangle$ and $\kbT$ stems from the similarity in size between the probe and the bath ALPs. 

\begin{figure}
  \centering
  \includegraphics[width=\linewidth]{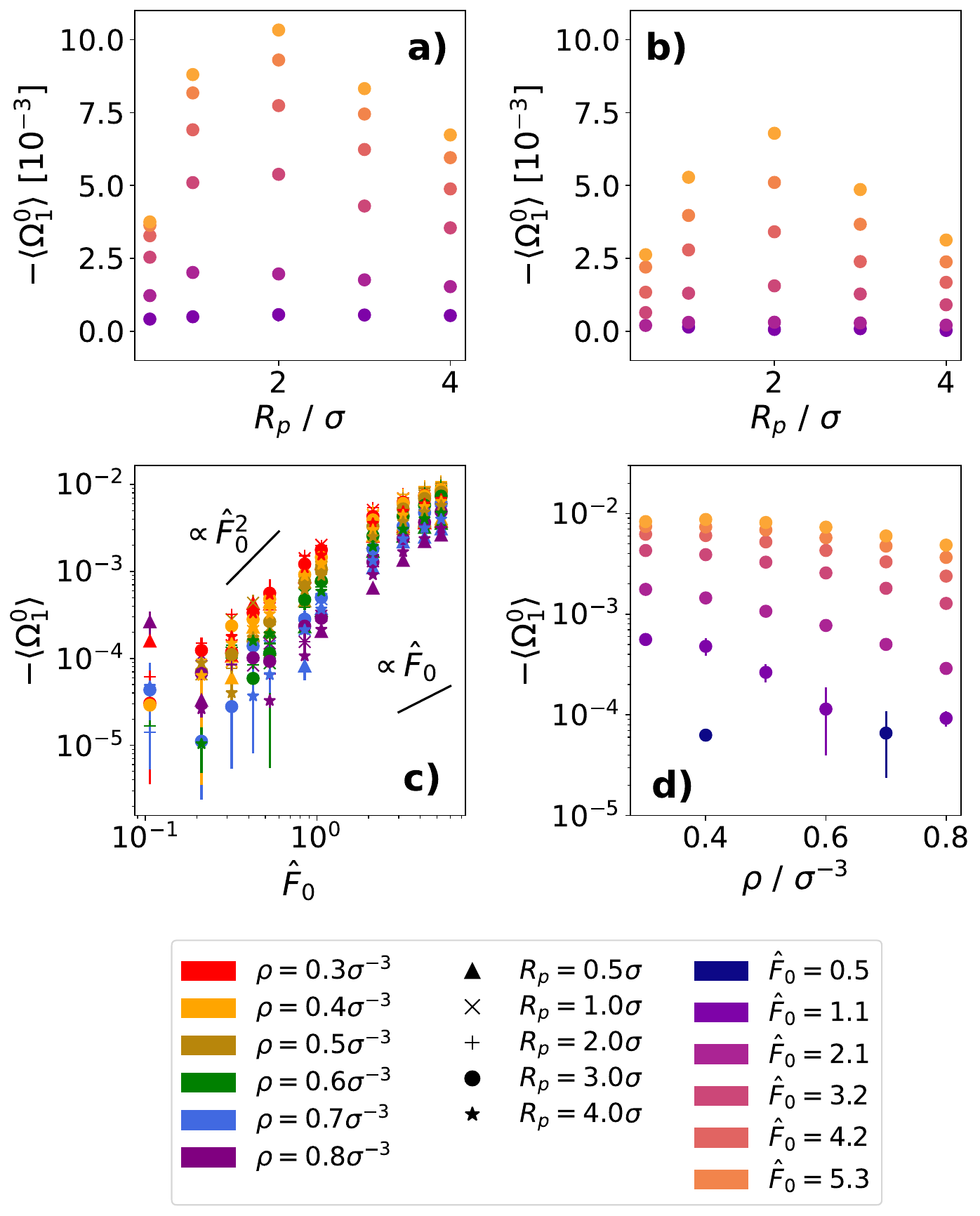}  
\caption{Average dipole moment, $\langle \Omega^0_1\rangle$, as a function of textbf{a)}/\textbf{b)} the radius $R_p$ for a bath of density \textbf{c)} $\rho=0.3\sigma^{-3}$, \textbf{d)} $\rho=0.8\sigma^{-3}$, \textbf{c)} the active force of the ALPs $\hat{F}_0$, \textbf{d)} the density $\rho$ for a probe of radius $R_p=3.0\sigma$.}
\label{fig:dip_comp}
\end{figure}

Fig.~\ref{fig:dip_temp}b shows the same data for probe particles with $R_p> 1.0 \sigma$ only. They roughly collapse onto one master curve, where  $\Delta k_B T^{\text{Probe}}_{\text{ALP bath}}$ initially increases linearly, and then quadratically as a function of $|\langle\Omega^0_1\rangle|$.
The solid black lines show the corresponding slopes for reference.
Overall, this positive correlation shows that the kinetic temperature (and hence the velocity of the probe, and probably active-like behavior as a whole) is mainly driven by the accumulation of particles behind it for larger probe sizes. Based on this positive correlation as well as the fact that the peak of the dipole moment exhibits non-monotonic behavior, we infer that the non-monotonicity of the kinetic temperature stems from the accumulation of particles around, and in particular behind, the probe. To investigate this further, we graph $(-\langle\Omega_1^0\rangle)$ as a function of $R_p$ in Figs.~\ref{fig:dip_comp}a) and b). We see that non-monotonic behavior indeed occurs for $\hf\gtrapprox2.1$, which matches the results of Fig.~\ref{fig:temps_rad} for the non-monotonic behavior of $\Delta \kbT_\mathrm{ALP\ Bath}^\mathrm{Probe}$ as a function of $R_p$. For active forces above this value, $|\langle\Omega_1^0\rangle|$ exhibits a maximum around $R_p \approx 2 \sigma$, which is reasonably 
close, but not identical, to the value $R_p \approx 3 \sigma$ where the maximum of the 
kinetic energy was observed.

We additionally plot $(-\langle\Omega_1^0\rangle)$ as a function of $\hat{F}_0$ and $\rho$ in Fig.~\ref{fig:dip_comp}c) and d) respectively. Fig.~\ref{fig:dip_comp}c) shows that $|\langle \Omega^0_1\rangle|$ scales approximately quadratically for low values of $\hf$ and approximately linearly for high values of $\hf$. This scaling is unsurprising given the scaling in Fig.~\ref{fig:dip_temp} and that $\Delta \kbT_\mathrm{ALP\ Bath}^\mathrm{Probe}\propto\hf^2$. The crossover from quadratic to linear occurs at $\hf\approx1$. At $\hat{F}_0=1$, the dominant force switches from the thermal forces to the ``pulling," active forces~\cite{ME}. We thus infer that this switch in the prominent forcing mechanism evokes the change in scaling.

In Fig.~\ref{fig:dip_comp}d), we see that $|\langle \Omega^0_1\rangle|$ decreases with increasing density. As the bath becomes more dense, crowding causes the distribution of ALPs around the probe to become more uniform. Therefore, the net dipole moment becomes less prominent.


\section{Conclusions and outlook}
\label{sec:conc_bi}

In sum, we have investigated the mechanism for the active-particle-like behavior of a probe immersed in an active bath. We have shown that this behavior as well as the enhanced kinetic temperature of a probe cannot simply be attributed to the convective motion of the active bath. The impenetrable boundary 
of the probe contributes significantly to these adopted dynamics. It causes active bath particles to accumulate behind the probe with respect to its instantaneous velocity. Once gathered behind the probe, particles are forced to move in the same direction, leading to correlations among bath particles with sufficiently high activity. This gathering of active bath particles pushes the probe, inducing transient positive feedback which can be seen in the memory kernel. This, in turn, promotes the active-particle-like behavior of the probe, in particular its enhanced kinetic temperature.

These boundary interactions, and consequently the dynamics of both the probe and the active fluid, are highly contingent on the specific configuration of the probe boundary. For probes with a radius approximately equal to that of the bath particles, the accumulation mechanism is significantly less effective. Furthermore, the kinetic temperature difference between the probe and the active fluid is dependent on the probe radius. The variability the boundary interactions furnishes us with another non-equilibrium signature of a probe immersed in an active bath: probes of different sizes acquire different kinetic temperatures, even when immersed in the same active bath.

We additionally found that the kinetic temperature difference between the immersed probe and the active bath scales non-monotonically with the probe size. This non-monotonicity is driven by the accumulation of active particles behind the probe. This accumulation must behave non-monotonically as a function of the probe size: for small probes, the surface area is too low to generate accumulation; for large probes, the probe 
velocity goes to zero, meaning the accumulation must be 
uniform.


Because our active particles are coupled to a Langevin thermostat, the hydrodynamic interactions in the fluid are effectively screened. In future work, it would be interesting to study systems with hydrodynamic interactions, which would certainly impact the correlations both within the active
fluid and between the active fluid and the passive probe.
Furthermore, given the importance of the boundary in both the dynamics of the probe and the active fluid, it would be interesting to consider a boundary with localized deformations. It has been shown that such boundary deformities can induce long-range effects in the bulk active fluid~\cite{Ben_Dor_2022}, which may in turn alter the dynamics of the immersed probe.

\section*{Acknowledgements}
This work was funded by the Deutsche Forschungsgemeinschaft (DFG) via Grant 233630050, TRR 146, Project A3. Computations were carried out on the Mogon Computing Cluster at ZDV Mainz. We thank Martin Hanke, Niklas Bockius, and Thomas Speck for useful discussions.

\bibliographystyle{unsrt}
\bibliography{refs.bib}

\appendix

\section{Velocity distribution for probes of different sizes}
\label{sec:app_veldist}
\label{sec:supp_calc_pv}


To calculate $P(\hv)$ from our simulation data, we calculate the absolute velocity, $|\mathbf{v}|$, of the particle for each time step. We then assign this value of $|\mathbf{v}|$ to an appropriate bin, each of length $dv$, to find the absolute velocity distribution $N[|\mathbf{v}|]$. We divide each of these bins by its true volume, $\delta V=\frac{4}{3} \pi\Big((v+\frac{dv}{2})^3-(v-\frac{dv}{2})^3\Big)$, and scale the distribution by a factor of $\sqrt{{m}/{\kbT}}$ to find $P(\hv)$. The distribution calculated from simulation data is normalized such that $\sum_{|\hv|} P(\hv)\delta V=1$. Theoretical distributions of $P(\hv)$ are normalized such that $\int_\infty\mathrm{d}\hat{v}~4\pi \hat{v}^2 P(\hv)=1$. 

\begin{figure}
  \centering
  \includegraphics[width=0.95\linewidth]{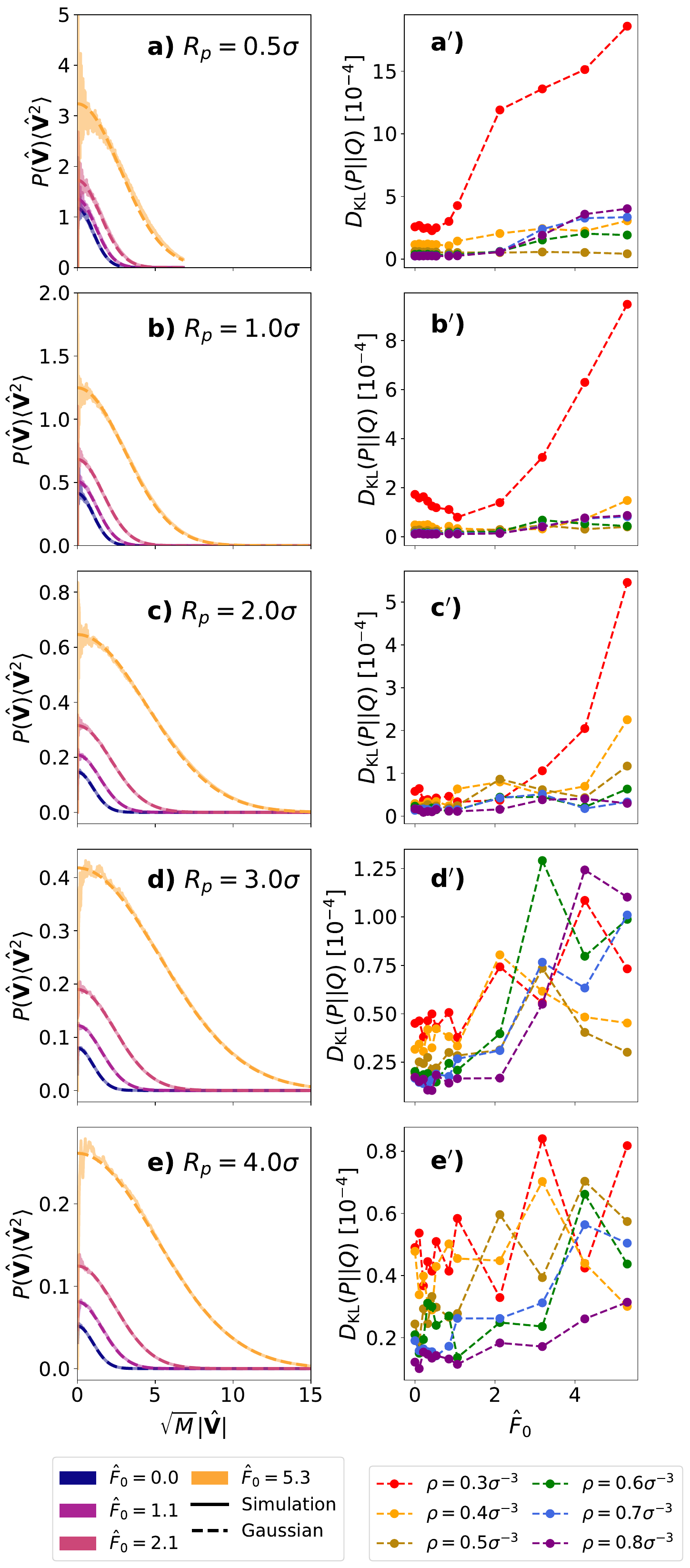}  
\caption{\textbf{a}-\textbf{e)}: Velocity distributions of different sized probes immersed in a bath of density $\rho=0.3\sigma^{-3}$ for different driving forces $\hat{F}_0$. Solid lines show simulation data, whereas dotted lines show zero-centered Gaussian distributions with the same standard deviation for comparison. The x-axis has been rescaled by $\sqrt{M}$ for better visibility and the x-axis has been rescaled by $\langle\hat{\mathbf{V}}^2\rangle$. \textbf{a$^{\prime}$}-\textbf{e$^{\prime}$)}: Corresponding relative entropies (Eq.~(\ref{eq:re})) between the simulation data and Gaussian distributions with the same standard deviation. Each row corresponds to a different sized probe from Table~\ref{tab:mvratio}: \textbf{a}/\textbf{a$^{\prime}$)} $R_p=0.5\sigma$, \textbf{b}/\textbf{b$^{\prime}$)} $R_p=1.0\sigma$, \textbf{c}/\textbf{c$^{\prime}$)} $R_p=2.0\sigma$, \textbf{d}/\textbf{d$^{\prime}$)} $R_p=3.0\sigma$, and \textbf{e}/\textbf{e$^{\prime}$)} $R_p=4.0\sigma$. Densities in the legend are expressed in LJ units of $\sigma^{-3}$.}
\label{fig:veldist_mvratio}
\end{figure}

In Figs.~\ref{fig:veldist_mvratio}a-e), we plot the velocity distributions $P(\hat{\mathbf{v}})$, of the different sized probes listed in Table~\ref{tab:mvratio} in a bath of density $\rho=0.3\sigma^{-3}$. We also show zero-centered Gaussian distributions with the same standard deviation as the simulation data. At least qualitatively, the Gaussian distributions seem to match the simulation data very well.

To better quantify this deviation from a Gaussian distribution, we calculate the relative entropy (Kullback-Leibler divergence) between the velocity distribution of the probe, as calculated from simulation data, and a zero-centered Gaussian distribution with the same standard deviation. The relative entropy between these two distributions is defined as:
\begin{equation}
\label{eq:re}
D_\mathrm{KL}(\mathcal{P}(\hv)||\mathcal{Q}(\hv))= \int_{\infty}\!\! \ud^3 \hat{v} \: \mathcal{P}(\hv)\ln{\Bigg(\frac{\mathcal{P}(\hv)}{\mathcal{Q}(\hv)}\Bigg)},
\end{equation}
where $\mathcal{P}(\hv)$ is the velocity distribution from simulation data and $\mathcal{Q}(\hv)$ is the reference Gaussian distribution.


This quantity is plotted in
Figs.~\ref{fig:veldist_mvratio}a$^{\prime}$-e$^{\prime}$). From Figs.~\ref{fig:veldist_mvratio}a$^{\prime}$-e$^{\prime}$), we see that the deviations from a Gaussian distribution are generally very small. We do see some relatively large deviations for small probes ($R_p\leq2.0\sigma$) immersed in active baths of low density ($\rho=0.3\sigma^{-3}$) in Figs.~\ref{fig:veldist_mvratio}a$^{\prime}$-c$^{\prime}$). However, when we look at the velocity distributions in Figs.~\ref{fig:veldist_mvratio}a-c), these deviations are only qualitatively noticeable near $|\mathbf{\hat{V}}|=0$, where the true volume ($\delta V$) goes to zero. Given that we divide by $\delta V$ to calculate $P(\hv)$, these deviations are likely due to numerical inaccuracies. We also see the deviations near $|\mathbf{\hat{V}}|=0$ for larger probes in Figs.~\ref{fig:veldist_mvratio}d) and e).

We attribute the larger values of relative entropy associated with smaller probes to poorer statistics for such systems: because the bath has a low density, the probe experiences fewer collisions with bath particles, which is only exacerbated by the small size of the probe. Therefore, we conclude that the deviations from a Gaussian distribution are due more to the less robust statistics in these systems than a fundamental lack of Gaussianity in the probe velocity distribution.


\section{Active Langevin particle persistence length}
\label{sec:app_pl}
\begin{figure}
  \centering
  \includegraphics[width=0.7\linewidth]{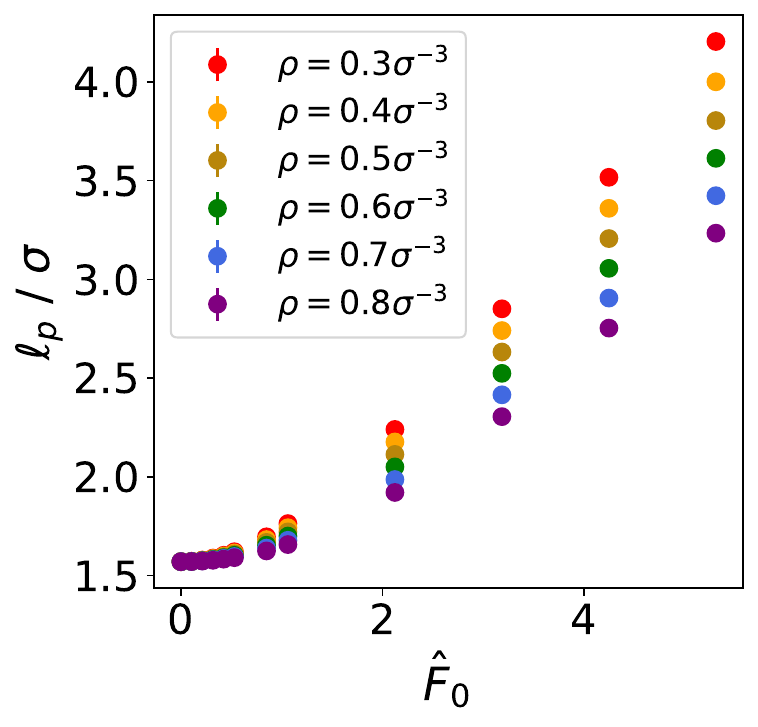}  
\caption{Persistence length of an ALP for different densities (shown as different colors) as a function of the active force $\hf$.}
\label{fig:supp_lp}
\end{figure}

The dynamics of active particles can generally be characterized by two parameters: the speed of propulsion and the rotational diffusion time. These two parameters can be used to define a persistence length during which an active particle travels without reorienting. For an active Brownian particle, this persistence length is typically defined as $\ell_p=v_0/((d-1)D_R)$, where $v_0$ is the propulsion velocity, $d$ is the dimensionality of the system, and $D_R$ is the rotational diffusion constant. In analogy with this definition, we define the persistence length of an ALP as:

\begin{equation}
    \ell_p=\frac{\sqrt{\langle\mathbf{v}^2\rangle}}{(d-1)D_R}.
    \label{eq:supp_lp}
\end{equation}
We show the ALP persistence length for different densities as a function of $\hf$ in Fig.~\ref{fig:supp_lp}.

From Fig.~\ref{fig:supp_lp}, we see that $\ell_p>R_p$ at all active forces for probes of $R_p\lesssim1.5$. This characteristic of the ALPs likely  contributes to the qualitatively different behavior we see for probes of radii $R_p=0.5\sigma$ and $R_p=1.0\sigma$. Furthermore, $\ell_p$ grows with the active force of the ALPs. In Section~\ref{sec:kin_temp_mvratio}, we see that the kinetic temperature difference between the probe and the bath ALPs exhibits non-monotonic behavior as a function of probe radius, reaching a maximum at some probe radius $R_\mathrm{max}$. We found in Section~\ref{sec:kin_temp_mvratio} that the value of $R_\mathrm{max}$ decreases with increasing bath activity (see Fig.~\ref{fig:spline_max}a)); therefore, given the opposite behavior of $\ell_p$ and $R_\mathrm{max}$ as function of $\hf$, we conclude that the persistence length of the ALPs cannot explain the non-monotonic behavior of the kinetic temperature difference.

\section{Stochastic force distribution}
\label{sec:app_sf}
From a trivial rewriting of Eq.~(\ref{eq:gle}), we calculate the stochastic force on the probe particle directly from our simulation data. We find that, for probes of all radii, the stochastic force distribution on the probe is Gaussian, as has previously been shown for a probe of radius $R_p=3.0\sigma$~\cite{ME}. However, it is notable that there are some deviations from a Gaussian distribution for probes with smaller radii. In Ref.~\cite{ME}, it has been shown that the stochastic force distribution of an immersed probe exhibits deviations from a Gaussian distribution at low densities, even when the bath is passive. The peaks in the simulation are sharper than the Gaussian distribution. Given this result, we attribute the deviations we see here 
to small numbers,
i.e. few collisions with bath particles, caused by the small size of the immersed probe, such that the central limit theorem no longer applies.


\begin{figure}
  \centering
  \includegraphics[width=\linewidth]{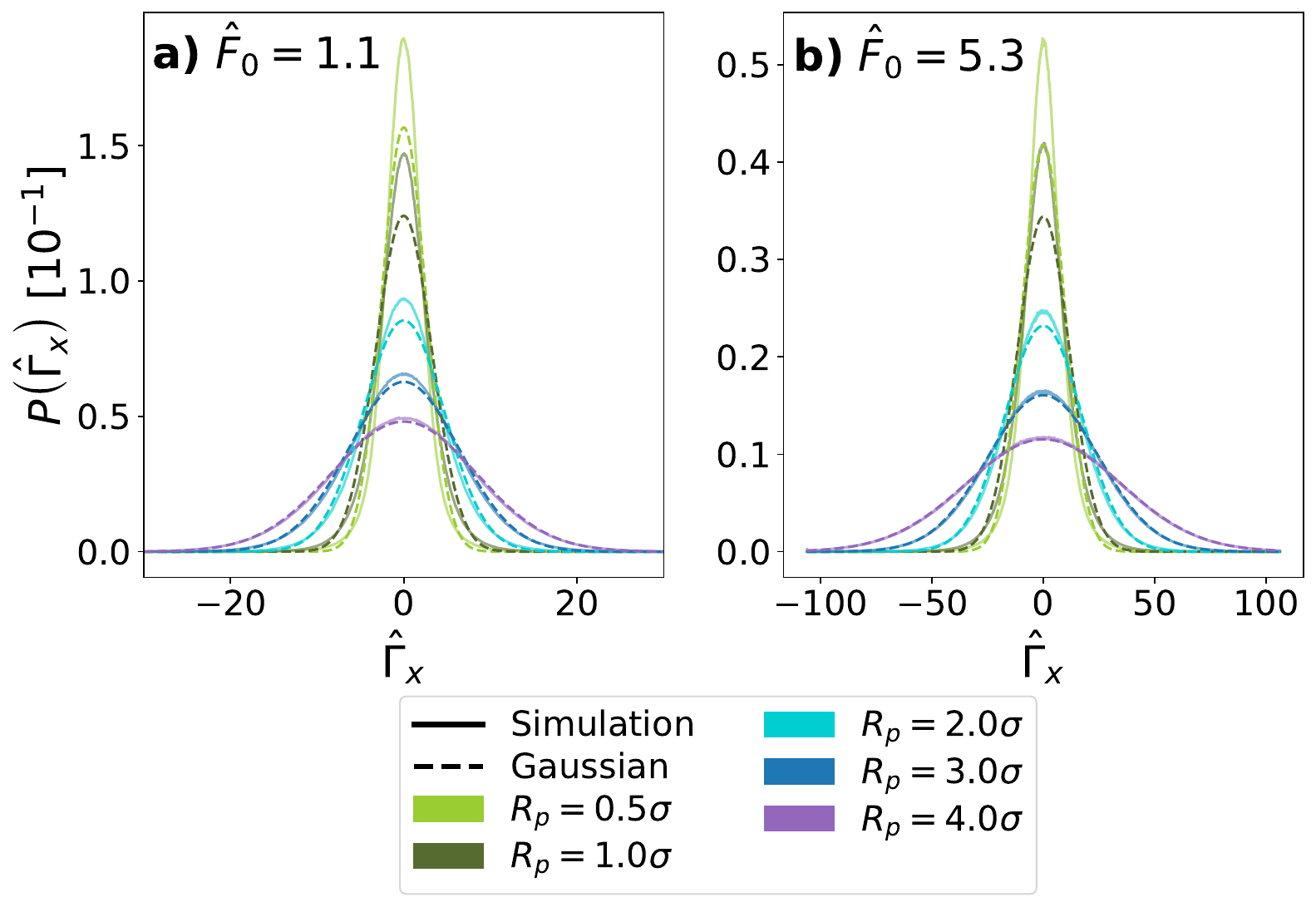}
\caption{Stochastic force distribution of different sized probes immersed in a bath of density $\rho=0.4\sigma^{-3}$. The solid lines show simulation data, whereas the dotted lines show zero-centered Gaussian distributions with the same standard deviation. \textbf{a)} The stochastic force distribution for a bath with activity $\hat{F}_0=1.1$. \textbf{b)} The stochastic force distribution for an active bath with activity $\hat{F}_0=5.3$.}
\label{fig:sfx}
\end{figure}



\section{Correlations and density in an active Langevin particle bath}
\label{sec:app_corr}

In Section~\ref{sec:kin_temp_mvratio}, we saw that the kinetic temperature difference between the probe and the bath ALPs exhibits non-monotonic behavior as a function of probe radius, reaching a maximum at some probe radius $R_\mathrm{max}$. We have already shown that the persistence length of the bath ALPs cannot explain this non-monotonic behavior in Appendix~\ref{sec:app_pl}. For baths of all densities and activities, we found the value of $R_\mathrm{max}\approx3\sigma$. Therefore, we would like to assess whether the length $3.0\sigma$ somehow characterizes the bath dynamics. Here, we specifically investigate correlations among ALPs.

\begin{figure}
  \centering
  \includegraphics[width=0.7\linewidth]{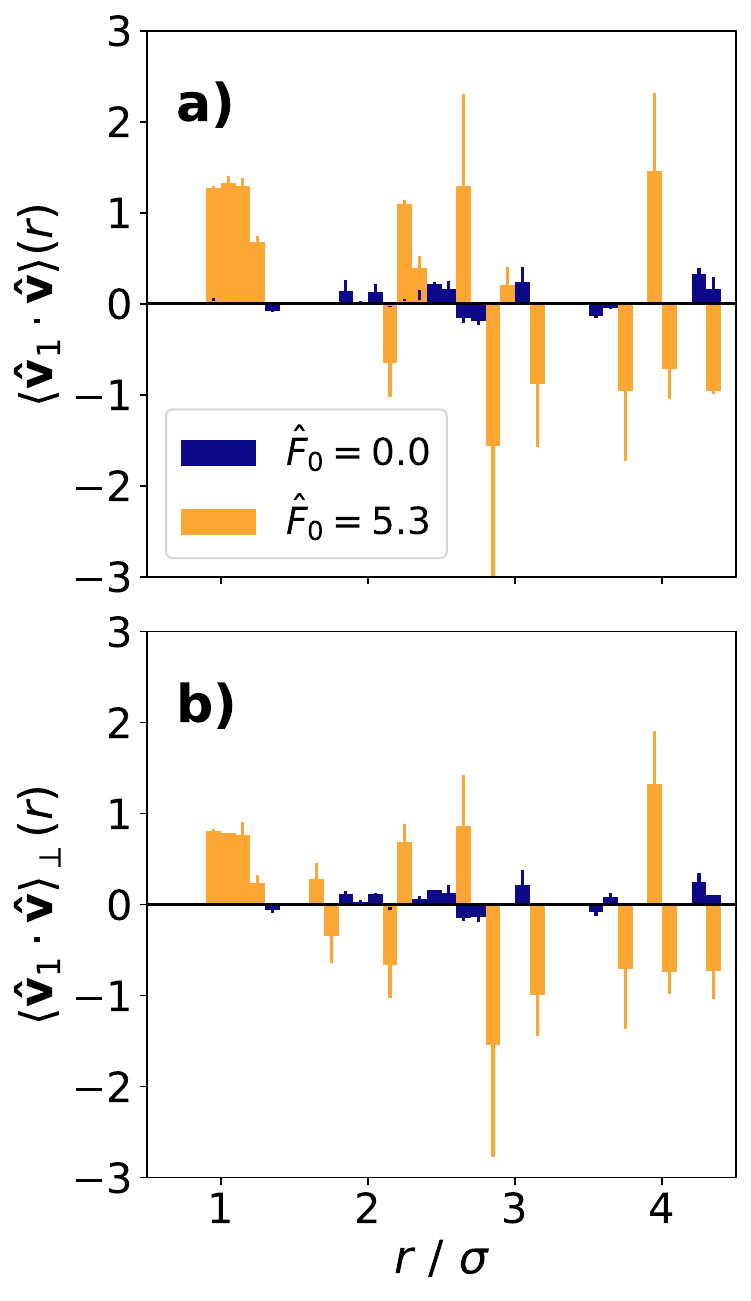}  
\caption{Velocity alignment among active Langevin particles in a bath of global density $\rho_0=0.8\sigma^{-3}$. \textbf{a)} Alignment of ALP velocities with that of a randomly chosen ALP ($\langle \mathbf{v}_1\cdot\mathbf{v}\rangle (r)$, see Eq.~\ref{eq:velcorr}). \textbf{b)} Alignment of ALP velocities located on the axis perpendicular to the velocity of a randomly chosen ALP ($\langle \mathbf{v}_1\cdot\mathbf{v}\rangle_\perp (r)$, see Eq.~\ref{eq:velcorr_perp}).}
\label{fig:app_alp_velcorr}
\end{figure}

For the purpose of this investigation, we examine a bath of ALPs (described by Eq.~\ref{eq:alp_int}), absent of the probe. For a single, randomly selected ALP, we then calculate the alignment of other ALPs in the vicinity using Eq.~\ref{eq:velcorr}. In Eq.~\ref{eq:velcorr}, we simply replace $\mathbf{V}$ with $\mathbf{v_1}$, the velocity of the randomly selected ALP (see Fig.~\ref{fig:app_alp_velcorr}a)). We are especially interested in correlations between $\mathbf{v_1}$ and the velocities of ALPs which are positioned on the axis perpendicular to $\mathbf{v_1}$. Correlations amongst these perpendicularly located particles would allow for coordinated pushing of the probe. Therefore, in Fig.~\ref{fig:app_alp_velcorr}b), we specifically examine the quantity:

\begin{equation}
\langle \mathbf{v}_1\cdot\mathbf{v}\rangle_\perp (r)=\frac{1}{\rho(r)}\sum_{i\in \delta \mathbf{r}}\mathbf{v}_1\cdot\mathbf{v}~\frac{|\mathbf{v}_1\times\mathbf{r}_i|}{|\mathbf{v}_1| |\mathbf{r}_i|},
\label{eq:velcorr_perp}
\end{equation}
where $\mathbf{r}_i$ is the distance vector from ALP $i$ to the selected ALP.

In agreement with our results in Section~\ref{sec:bubb}, we see that the motion of particles in a passive bath is uncorrelated. In an active bath, we see that the velocities of the ALPs do become correlated, again in agreement with Chapter~\ref{sec:bubb}. We find these correlations up to a length of $\sim1.0\sigma$. A significant amount of these correlations are among particles positioned along the axis perpendicular to $\mathbf{v}_1$ (see Fig.~\ref{fig:app_alp_velcorr}b)). On larger length scales, the ALPs become uncorrelated again. The correlation length scale of $\sim1.0\sigma$ does not match the value of $R_\mathrm{max}\approx3\sigma$ which we saw in Chapter~\ref{sec:kin_temp_mvratio}. Therefore, these correlations cannot explain such a value of $R_\mathrm{max}$.

\end{document}